\documentclass{ws-ijmpa}

\newcommand{\be}[1]{\begin{equation}\label{#1}}
\newcommand{\beq}{\begin{equation}}
\newcommand{\ee}{\end{equation}}
\newcommand{\beqn}[1]{\begin{eqnarray}\label{#1}}
\newcommand{\eeqn}{\end{eqnarray}}

\newcommand{\dub}[2]{\left(\begin{array}{c}{#1}\\{#2}
\end{array}\right)}
\newcommand{\dubs}[2]{\begin{array}{l}{#1}\\{#2} \end{array}}
\newcommand{\mat}[4]{\left(\begin{array}{cc}{#1}&{#2}\\{#3}&{#4}
\end{array}\right)}

\renewcommand{\to}{\rightarrow}

\def\ov{\overline}

\def\lsim{\raise0.3ex\hbox{$\;<$\kern-0.75em\raise-1.1ex
\hbox{$\sim\;$}}}
\def\gsim{\raise0.3ex\hbox{$\;>$\kern-0.75em\raise-1.1ex
\hbox{$\sim\;$}}}

\def\bg{{\bf g}}

\def\cG{{\cal G}}

\def\cL{{\cal L}}
\def\cM{{\cal M}}

\def\al{\alpha}
\def\ga{\gamma}
\def\Ga{\Gamma}

\def\la{\lambda}
\def\La{\Lambda}  

\def\Om{\Omega}
\def\eps{\varepsilon}

\def\DN{\Delta N_\nu}

\def\rhb{\bar{\rho}}

\def\bg{\bar{g}_\ast}

\def\tpsi{\tilde{\psi}}
\def\tphi{\tilde{\phi}}

\def\tq{\tilde{q}}
\def\tl{\tilde{l}}

\def\tR{\tilde{R}}
\def\tN{\tilde{N}}
\def\te{\tilde{e}}
\def\tu{\tilde{u}}
\def\td{\tilde{d}}
\def\tnu{\tilde{\nu}}

\def\lpr{l^\prime}
\def\phpr{\phi^\prime}

\def\Apr{A^\prime}

\def\barl{\bar{l}}
\def\barphi{\bar{\phi}}
\def\barlpr{\bar{l}^\prime}
\def\barphpr{\bar{\phi}^\prime}

\begin{document}

\markboth{Zurab Berezhiani}{Mirror World ...}

\catchline{}{}{}{}{}

\title{
%CONCEPT OF 
MIRROR WORLD AND ITS COSMOLOGICAL CONSEQUENCES} 

\author{ZURAB BEREZHIANI}
\address{
%AUTHOR ADDRESS AND EMAIL \\
 Dipartimento di Fisica, Universit\`a di L'Aquila, 67010 Coppito,
L'Aquila, and \\
INFN, Laboratori Nazionali del Gran Sasso, 67010 Assergi, L'Aquila,
Italy; \\
e-mail: berezhiani@fe.infn.it
}

%\author{SECOND AUTHOR'S NAME}
%\address{AUTHOR ADDRESS AND EMAIL}  

%%%%%%%%%%%%%%%%%%%%%%%%%%%%%%%%%%%%%%%%%%%%%%%%%%%%%%%%%%%%
% You may repeat \author \address as often as necessary    %
%%%%%%%%%%%%%%%%%%%%%%%%%%%%%%%%%%%%%%%%%%%%%%%%%%%%%%%%%%%%

\maketitle

%\begin{history}
%\received{(DAY MONTH YEAR)}
%\revised{(DAY MONTH YEAR)}
%\end{history}

\begin{abstract}
We briefly review the concept of a parallel `mirror' world
which has the same particle physics as the observable world
and couples to the latter by gravity and perhaps other very 
weak forces. 
The nucleosynthesis bounds demand that the mirror world
should have a smaller temperature than the ordinary one.
By this reason its evolution should substantially deviate 
from the standard cosmology as far as the crucial epochs like
baryogenesis, nucleosynthesis etc. are concerned.
In particular, we show that in the context of certain 
baryogenesis scenarios, the baryon asymmetry in the mirror 
world should be larger than in the observable one. 
Moreover, we show that mirror baryons could naturally 
constitute the dominant dark matter component of the Universe, 
and discuss its cosmological implications.  
%baryons representing a kind of self-interacting dark matter
%for the large scale structure formation, the CMB anysotropy,
%the galactic halo structures, microlensing, 
%etc. are briefly discussed.
\end{abstract}

% TYPE KEYWORDS HERE, SEPARATE THEM WITH SEMI-COLON 

%%%%%%%%%%%%%%%%%%%%%%%%%%%%%%%%%%%%%%%%%%%%%%%%%%%%%%%%%%%%
% The main text of your paper   begins here                          %
%%%%%%%%%%%%%%%%%%%%%%%%%%%%%%%%%%%%%%%%%%%%%%%%%%%%%%%%%%%%

\section{Introduction} 

%$l$ ~ ${\cal l}$    ~  $q$ ~ ${\cal q}$

%$\bYu$  ~~~~  ~~~~~ $\bY_u$  ~~~ $Y_u$ ~~~~ 

The old idea that there can exist a hidden mirror sector of
particles and interactions which is the exact duplicate
of our visible world \cite{LY}
%and communicates the later through the gravity
%and perhaps also via some other messengers,
%Various physical and astrophysical implications
%have been studied in several subsequent papers
has attracted a significant interest over the last years. 
The basic concept is to have a theory given by the
product $G\times G'$ of two identical gauge factors with
the identical particle contents, which could naturally emerge
e.g. in the context of $E_8\times E'_8$ superstring.

In particular, one can consider
a minimal symmetry $G_{\rm SM}\times G'_{\rm SM}$, 
where $G_{\rm SM}=SU(3)\times  SU(2)\times U(1)$ stands for the  
standard model of observable particles: three families of
quarks and leptons 
%$q_i, ~u^c_i, ~d^c_i; ~l_i, ~e^c_i$ ($i=1,2,3$) 
and the Higgs, 
%$\phi$,
while $G'_{\rm SM}=[SU(3)\times SU(2)\times U(1)]'$ is its mirror
gauge counterpart with analogous particle content:
three families of mirror quarks and leptons 
%fermions $q'_i, ~u'^c_i, ~d'^c_i; ~l'_i, ~e'^c_i$  and 
and the mirror Higgs. 
%$\phi'$.
(From now on all fields and quantities of the
mirror (M) sector will be marked by $'$ 
to distinguish from the ones belonging to the
observable or ordinary (O) world.)
The M-particles are singlets of $G_{\rm SM}$ and vice versa, 
the O-particles are singlets of $G'_{\rm SM}$. 
Besides the gravity, the two sectors could communicate 
by other means. In particular, ordinary photons 
could have kinetic mixing with mirror photons 
\cite{Holdom,Glashow86,Glashow87},  
ordinary (active) neutrinos could mix with mirror 
(sterile) neutrinos \cite{FV,BM}, or
two sectors could have a common 
gauge symmetry of flavour \cite{PLB98}. 

A discrete symmetry $P(G\leftrightarrow G')$ interchanging
corresponding fields of $G$ and $G'$, so called mirror parity,
guarantees that both particle sectors are described by the
same Lagrangians, with all coupling constants 
(gauge, Yukawa, Higgs) having the same pattern, and thus
their microphysics is the same. 

If the mirror sector exists, then the Universe  
along with the ordinary photons, neutrinos, baryons, etc. 
should contain their mirror partners.  
One could naively think that due to mirror parity the
ordinary and mirror particles should have the same cosmological 
abundances and hence the O- and M-sectors should have the same
cosmological evolution. 
However, this would be in the immediate conflict 
with the Big Bang nucleosynthesis (BBN) bounds
on  the effective number of extra light neutrinos,
since the mirror photons, electrons and neutrinos
would give a contribution to the Hubble expansion rate 
equivalent to $\DN\simeq 6.14$.
Therefore, in the early Universe the M-system should have 
a lower temperature than ordinary particles. 
This situation is plausible if the following conditions  
are satisfied:

A. After the Big Bang the two systems are born with different 
temperatures, namely the post-inflationary reheating temperature 
in the M-sector is lower than in the visible one,
$T'_R < T_R$. This can be naturally achieved in certain models
\cite{KST,BDM,BV}.

B. The two systems interact very weakly,  
so that they do not come into thermal equilibrium
with each other after reheating.
This condition is automatically fulfilled  if the two worlds
communicate only via gravity.
If there are some other effective couplings 
between the O- and M- particles, they 
have to be properly suppressed. 
%suppressed by a large mass factor $M > T_R$ or so.  
%$M\sim M_{P}$ or so.  

C. Both systems expand adiabatically, without significant 
entropy production. 
%e.g. due to the first order electroweak or QCD phase transitions 
%which could heat the M-sector and increase its temperature 
%to the observable one. 
If the two sectors have different reheating temperatures, 
during the expansion of the Universe they evolve independently
and their temperatures remain different at later stages, 
$T' < T$, 
then the presence of the M-sector would not affect primordial 
nucleosynthesis in the ordinary world.  

At present, the temperature of ordinary relic photons 
is  $T\approx 2.75$ K, and the mass density of ordinary 
baryons constitutes about $5\%$ of the critical 
density. 
Mirror photons should have smaller temperature $T' < T$, 
so their number density is $n'_\ga = x^3 n_\ga$, where    
$x=T'/T$. This ratio is a key parameter in our 
further considerations since it remains nearly invariant 
during the expansion of the Universe. 
The BBN bound on $\DN$ implies the upper bound  
$x < 0.64\, \DN^{1/4}$. 
As for mirror baryons, {\it ad hoc} 
their number density $n'_b$  can be larger than $n_b$, 
and if the ratio $\beta = n'_b/n_b$ is about 5 or so, 
they could constitute the dark matter of the Universe.  
%or at least its significant fraction.  
 
In this paper we study the cosmology of the mirror 
sector and discuss the comparative time history      
of the two sectors in the early Universe.
We show that due to the temperature difference, in the
mirror sector all key epochs as the baryogenesis,
nucleosynthesis, etc. proceed at somewhat different
conditions than in the observable Universe.
In particular, we show that in certain baryogenesis scenarios 
the M-world gets a larger baryon asymmetry than the O-sector, 
and it is pretty plausible that $\beta > 1$.\cite{BCV} 
This situation emerges in a particularly appealing way 
in the leptogenesis scenario due to the lepton number  
leaking from the O- to the M-sector which leads to  
$n'_b \geq n_b$, and can thus explain
the near coincidence of visible and dark components  
in a rather natural way \cite{BB-PRL,Bamberg}.

\section{Mirror world and mirror symmetry}

\subsection{Particles and couplings in the ordinary world}

Nowadays almost every particle physicist knows that 
particle physics is described by the Standard Model (SM) 
based on the gauge symmetry 
$G_{\rm SM} =SU(3)\times  SU(2)\times U(1)$, 
which has a chiral fermion pattern: 
%in the limit of unbroken gauge symmetry 
fermions are represented as Weyl spinors, so that 
the left-handed (L) quarks and leptons $\psi_L = q_L,l_L$ 
and right-handed (R) ones $\psi_R = q_R,l_R$
transform differently under the $SU(2)\times U(1)$ gauge factor. 
More precisely, the fermion content is the following: 
\beqn{SM-L} 
&
l_L = \dub{\nu_L}{e_L} \sim (1,2,-1) ; ~~~~ 
&
l_R = \left\{ \dubs { N_R \sim (1,1,0)~~~(?) } 
{ e_R \sim (1,1,-2) } \right. 
\nonumber \\ 
&
q_L = \dub{u_L}{d_L} \sim (3,2,1/3) ; ~~~~ 
&
q_R= \left\{\dubs { u_R \sim (3,1,4/3) } 
{ d_R \sim (3,1,-2/3) } \right. ,  
\eeqn
where the brackets explicitly indicate the  
$SU(3)$ and $SU(2)$ content of the multiplets and their 
$U(1)$ hypercharges. In addition, one prescribes 
a global lepton charge $L=1$ to the leptons $l_L,l_R$ and 
a baryon charge $B=1/3$ to quarks $q_L,q_R$, so that baryons 
consisting of three quarks have $B=1$. 

The $SU(2)\times U(1)$ symmetry is spontaneously broken 
at the scale $v=174$ GeV and $W^\pm,Z$ gauge 
bosons become massive. At the same time 
charged fermions get masses via the Yukawa couplings
($i,j=1,2,3$ are the fermion generation indexes) 
\be{Yuk-LR}
{\cal L}_{\rm Yuk} = 
\,Y^u_{ij}\, \ov{u_R}_i\, q_{Lj} \, \phi_u\, +\, 
Y^d_{ij}\, \ov{d_R}\, q_{Lj}\, \phi_d\, + \,
Y^e_{ij}\, \ov{e_R}_i\, l_{Lj}\, \phi_d \, + \, {\rm h.c.} 
%
%(Y^{u\ast}_{ij} \tq_i u_j \tilde{\phi}_2 +
%Y^{d\ast}_{ij} \tq_i d_j \tilde{\phi}_1 + 
%Y^{e\ast}_{ij} \tl_i e_j \tilde{\phi}_1)
\ee
where the Higgs doublets $\phi_u =\phi \sim (1,2,1)$ and 
$\phi_d = \tphi \sim (1,2,-1)$, in the minimal SM, 
are simply conjugated to each other: $\phi_d \sim \phi_u^\ast$. 
However, in the extensions of the SM, 
and in particular, in its supersymmetric version, 
$\phi_{u,d}$ are independent ("up" and "down") Higgs doublets 
with different vacuum expectation values (VEV) 
$\langle\phi_u\rangle = v_u$ and  $\langle\phi_d\rangle = v_d$, 
where $v_u^2 + v_d^2=v^2$, and their ratio is parametrized as 
$\tan\beta = v_u/v_d$.  

Obviously, with the same rights the Standard Model could be
formulated in terms of the field operators 
$\tpsi_R = C\ga_0\psi_L^\ast$ and $\tpsi_L = C\ga_0\psi_R^\ast$, 
%$\tpsi_R = C\ov{\psi_L}^T$ and $\tpsi_L = C\ov{\psi_R}^T$, 
where $C$ is the charge conjugation matrix. 
These operators now describe antiparticles which have 
opposite gauge charges  
%with respect to $G_{\rm SM} =SU(3)\times  SU(2)\times U(1)$:
as well as  opposite chirality than particles: 
%$\psi_L \to \tpsi_R$ and $\psi_R \to \tpsi_L$, 
\beqn{SM-R} 
&
\tl_R = \dub{\tnu_R}{\te_R} \sim (1,\bar2,1) ; ~~~~~~ 
&
\tl_L = \left\{ \dubs { \tN_L \sim (1,1,0)~~~(?) } 
{ \te_L \sim (1,1,2) } \right. 
\nonumber \\ 
&
\tq_R = \dub{\tu_R}{\td_R} \sim (\bar3,\bar2,-1/3) ; ~~~~ 
&
\tq_L = \left\{\dubs {\tu_L \sim (\bar3,1,-4/3) } 
{ \td_L \sim (\bar3,1,2/3) } \right. . 
\eeqn
Clearly, now antileptons $\tl_R,\tl_L$ have $L=-1$ and 
antiquarks $\tq_R,\tq_L$ have $B=-1/3$. 
The full system of particles and antiparticles 
can be presented in a rather symmetric way as: 
\be{LR} 
{\rm fermions}: ~~ \psi_L,~ \psi_R \, ;  ~~~
{\rm antifermions}: ~~  \tpsi_R,~ \tpsi_L .     
\ee 
However, one can simply redefine the notion of particles, 
namely, to call $\psi_L,\tpsi_L$ as L-particles and 
$\tpsi_R,\psi_R$, as R antiparticles 
%which are $C$-conjugated to 
\be{LL} 
L-{\rm fermions}: ~~ \psi_L,~ \tpsi_L \, ; ~~~ 
\tR{\rm -antifermions}: ~~ \tpsi_R,~ \psi_R    
%{\rm L-fermions}: ~~ F = \{ \psi_L,~ \tpsi_L \}\, ; ~~~ 
%{\rm R-antifermions}: ~~ \tilde{F} = \{ \tpsi_R,~ \psi_R \} . 
\ee 
%so that $\tF = C\ov{F}^T= C\ga_0F^\ast$. 
Hence, the standard model fields, including Higgses, 
can be recasted as:\footnote{
In the context of $N=1$ supersymmetry, the fermion as well as 
Higgs fields all become chiral superfields, and formally 
they can be distinguished only by matter parity $Z_2$ 
under which the fermion superfields change the sign while 
the Higgses remain invariant. 
Therefore, odd powers of what we call fermion 
superfields are excluded from the superpotential. 
} 
\be{LL-SM}
L{\rm -set}: ~~ (q,l,\tu,\td,\te,\tN)_L, ~\phi_u,\phi_d \, ; ~~~ 
\tR{\rm -set}: ~~ (\tq,\tl,u,d,e,N)_R, ~ \tphi_u,\tphi_d  
\ee
where $\tphi_{u,d}=\phi_{u,d}^\ast$, and the 
the Yukawa Lagrangian (\ref{Yuk-LR}) can be rewritten as 
\be{Yuk-LL}
{\cal L}_{\rm Yuk} = 
\tu^T Y_u q \phi_u + \td^T Y_d q \phi_d + \te^T Y_e l \phi_d  
\, + \, {\rm h.c.} 
%(u^T Y_u^\ast \tq \phi_u^\ast + d^T Y_d^\ast \tq \phi_d^\ast + 
%e Y_e^\ast \tl\phi_d^\ast) 
\ee
where the $C$-matrix  as well as the family indices are omitted 
for simplicity. 

In the absence of right-handed singlets $N$ 
there are no renormalizable Yukawa couplings which could 
generate neutrino masses. However, once the higher order 
terms are allowed, the neutrinos could get Majorana 
masses via the $D=5$ operators: 
\be{op-nu}
\frac{A_{ij}}{2 M} (l_i \phi_2) (l_j \phi_2) 
~ + ~ {\rm h.c.}
%\frac{A^\ast_{ij}}{2 M} (\tl_i \tphi_2) (\tl_j \tphi_2) 
\ee 
where $M$ is some large cutoff scale. 
% e.g. $M\sim M_{Pl}$  \cite{Weinb}. 
On the other hand, introducing the $N$ states is nothing 
but a natural way to generate effective operators (\ref{op-nu}) 
from the renormalizable couplings in the context of the seesaw 
mechanism.
Namely, $N$'s are gauge singlets and thus  
are allowed to have the Majorana mass terms 
$\frac12 (M_{ij} N_i N_j + M_{ij}^\ast \tN_i \tN_j)$. 
It is convenient to parametrize their mass matrix as 
$M_{ij} = G_{ij}M$, where $M$ is a typical mass scale 
and $G$ is a matrix of dimensionless Yukawa-like constants.  
On the other hand, the $\tN$ states can couple to $l$  
via Yukawa terms analogous to (\ref{Yuk-LL}), and 
thus the whole set of Yukawa terms obtain the pattern: 
\be{Yuk-nus} 
l^T Y\tN \phi_u + \frac{M}{2} \tN^T G \tN + {\rm h.c.} 
%Y^{\nu\ast}_{ij} \tl_i N_j \tilde{\phi}_1)
\ee 
As a result, the effective operator (\ref{op-nu}) emerges after 
integration out the heavy states $N$ with $A = Y G^{-1}Y^T$.      
This makes clear why the neutrinos masses are small -- 
they appear in second order of the Higgs field $\phi$, cutoff 
by large mass scale $M$, $m_\nu \sim v^2/M$, 
while the charged fermion mass terms are linear in $v$. 
 
Obviously, the redefinition (\ref{LL-SM}) is very convenient 
for the extension of the Standard Model related 
to supersymmetry and grand unified theories. 
For example, in $SU(5)$ model the L fermions fill the 
representations $f(\td,l)\sim \bar5$ and 
$t(\tu,q,\te)\sim 10$, while 
the R antifermions are presented by $\bar{f}(d,\tl)\sim 5$ 
and $\bar{t}(u,\tq,e)\sim \ov{10}$. 
In $SO(10)$ model all L fermions 
in (\ref{LL}) sit in one representation 
$L\sim 16$, while the R anti-fermions sit in 
$\tR \sim \ov{16}$. 

The Standard Model crudely violates one of the possible 
fundamental symmetries of the Nature, parity, since its 
particle content and hence its Lagrangian is not symmetric with 
respect to exchange of the L particles to the R ones. 
In particular, the gauge bosons of $SU(2)$ couple to the $\psi_L$ 
fields but do not couple to $\psi_R$.   
In fact, in the limit of unbroken $SU(2)\times U(1)$ symmetry, 
$\psi_L$ and $\psi_R$ are essentially independent particles 
with different quantum numbers. 
The only reason why we call e.g. states $e_L\subset l_L$ 
and $e_R$ respectively the left- and right-handed electrons, 
is that after the electroweak breaking down to $U(1)_{\rm em}$ 
these two have the same electric charges and form a massive 
Dirac fermion $\psi_e = e_L+e_R$. 

%since 
%CP-invariance is broken if the Yukawa constants are complex. 

Now what we call particles (\ref{SM-L}), 
the weak interactions are left-handed ($V-A$) 
since only the L-states couple to the $SU(2)$ gauge bosons. 
In terms of antiparticles (\ref{SM-R}), the weak interactions
would be right-handed ($V+A$), 
since now only R states couple to the $SU(2)$ bosons. 
Clearly, one could always redefine the notion of particles 
and antiparticles, to rename particles as antiparticles and 
vice versa. 
Clearly, the natural choice for what to call particles 
is given by the content of matter in our Universe.  
Matter, at least in our galaxy and its neighbourhoods, 
consists of baryons $q$ while 
antibaryons $\tq$ can be met only in accelerators or perhaps 
in cosmic rays. However, if by chance we would live in the 
antibaryonic island of the Universe, we would claim that 
our weak interactions are right-handed.  

In the context of the SM or its grand unified extensions,  
the only good symmetry between the left and right could be 
the CP symmetry between L-particles and R-antiparticles.  
E.g., the Yukawa couplings (\ref{Yuk-LL}) in explicit form 
read 
\be{Yuk-LLRR}
{\cal L} = 
(\tu^T Y_u q \phi_u + \td^T Y_d q \phi_d + \te^T Y_e l \phi_d)_L 
\, + \, 
(u^T Y_u^\ast \tq \tphi_u  + d^T Y_d^\ast \tq \tphi_d + 
e^T Y_e^\ast \tl\tphi_d)_R 
\ee
However, although these terms are written in an symmetric manner 
in terms of the $L$-particles and $\tR$-antiparticles, 
they are not invariant under $L\to \tR$ due to irremovable 
complex phases in the Yukawa coupling matrices.  
Hence, Nature does not respect the symmetry between particles and 
antiparticles, but rather applies the principle that 
"the only good discrete symmetry is a broken symmetry".  

It is a philosophical question, who and 
how has prepared our Universe at the initial state  
to provide an excess of baryons over antibaryons, 
and therefore fixed a priority of the $V-A$ form 
of the weak interactions over the $V+A$ one. 
It is appealing to think that the 
baryon asymmetry itself emerges due to the  
CP-violating features in the particle interactions, 
%of the Standard model or its extensions, 
and it is related to some fundamental physics beyond the 
Standard Model which is responsible for the primordial 
baryogenesis.

\subsection{Particles and couplings in the O- and M-worlds}

Let us assume now that there exists a mirror sector which 
has the same gauge group and the same particle content 
%and the same pattern of the coupling constants  
as the ordinary one. In the minimal version, when the 
O-sector is described by the gauge symmetry  
$G_{\rm SM} =SU(3)\times  SU(2)\times U(1)$ 
with the observable fermions and Higgses (\ref{LL-SM}), 
%three families of quarks and leptons 
%$q_i, ~\tu_i, \td_i; ~l_i, ~\te_i$ ($i=1,2,3$)
%and the Higgs doublet(s) $\phi_{u,d}$,
the M- sector would be given by the gauge group 
$G^\prime_{\rm SM} 
 = SU(3)^\prime \times SU(2)^\prime \times U(1)^\prime$
with the analogous particle content:
\be{LL-SMpr}
L'{\rm -set}: ~ (q',l',\tu',\td',\te',\tN')_L, 
~\phi'_u,\phi'_d \, ; 
~~~ 
\tR'{\rm -set}: ~ (\tq',\tl',u',d',e',N')_R, ~ \tphi'_u,\tphi'_d  
\ee

In more general view, one can consider a supersymmetric 
theory with a gauge symmetry $G\times G'$ based on 
grand unification groups as $SU(5)\times SU(5)^\prime$,
$SO(10)\times SO(10)^\prime$, etc.   
The gauge factor $G$ of the O-sector contains  
the vector gauge superfields $V$, 
and left chiral matter (fermion and Higgs) superfields $L_a$ 
in certain representations of $G$, while 
$G'$ stands for the M-sector with the gauge superfields $V'$, 
and left chiral matter superfields $L'_a\sim r_a$
in analogous representations of $G'$. 

The Lagrangian has a form $\cL_{\rm gauge} + \cL_{\rm mat}$. 
The matter Lagrangian is determined by 
the form of the superpotential which is a holomorphic 
function of L superfields and in general 
it can contain any gauge invariant combination of the latter: 
\be{super}
W = (M_{ab} L_aL_b + g_{abc}L_a L_b L_c)~ + ~
(M'_{ab} L'_aL'_b + g'_{abc}L'_a L'_b L'_c) + ... 
%\frac{h_{abcd}}{M} L_a L_b L_c L_d + ... 
\ee
where dots stand for possible higher order terms.  
Namely, we have 
\be{susy-W} 
\cL_{\rm mat} =  \int d\theta^2 W(L) ~ + ~ {\rm h.c.}  
\ee 
or, in explicit form, 
\beqn{susy} 
\cL_{\rm mat} & = & \int d\theta^2 
(M_{ab} L_aL_b + g_{abc}L_a L_b L_c + 
M'_{ab} L'_aL'_b + g'_{abc}L'_a L'_b L'_c )
\nonumber \\ 
& + & \int d\ov{\theta}^2 
(M^\ast_{ab}\tR_a\tR_b + g^\ast_{abc}\tR_a \tR_b \tR_c + 
M^{\prime\ast}_{ab}\tR'_a\tR'_b + g^{\prime\ast}_{abc}
\tR'_a \tR'_b \tR'_c)
\eeqn 
where $\theta$ and $\bar{\theta}$ are the Grassmanian 
coordinates respectively in $(1/2,0)$ and $(0,1/2)$ 
representations of the Lorentz group, and  
$\tR(\tR')=L^\ast(L^{\prime\ast})$ are 
the CP-conjugated right handed superfields.   
%in anti-representations $\bar{r}_a$ of $G$. 
%and contain states of R-set in (\ref{LL-SM}). 

One can impose a discrete symmetry between
these gauge sectors in two ways:    

A. Left-left symmetry $D_{LL}$:  
%which transforms superfields between O- and M-sectors as  
\be{DLL}
L_a \to L^\prime_a  ~~~(\tR_a \to \tR^\prime_a), ~~~ 
V \to V^\prime.
\ee
which for the coupling constants implies 
\be{DLL1}
M^\prime_{ab} = M_{ab}, ~~~~ g'_{abc} = g_{abc}    
\ee
Clearly, this is nothing but direct doubling 
and in this case the M-sector is an identical copy of the O-sector. 

B. Left-right symmetry $P_{LR}$: 
\be{PLR}
L_a \to \tR^\prime_a  ~~~(\tR_a \to L^\prime_a), ~~~ 
V \to V^\prime.
\ee
%Clearly, this is nothing but direct doubling 
which requires that 
\be{PLR1}
M^\prime_{ab} = M^\ast_{ab}, ~~~~ g'_{abc} = g^\ast_{abc}  
\ee
In this case the M-sector is a mirror copy of the 
ordinary sector, and this symmetry can be considered 
as a generalization of parity.\footnote{  
Obviously, if one imposes both $D_{LL}$ and $P_{LR}$, 
one would get CP-invariance as a consequence.}  

Either type of parity implies that the two sectors 
have the same particle physics.\footnote{ The mirror parity
could be spontaneously broken and the weak interaction scales
$\langle \phi \rangle =v$ and $\langle \phi' \rangle =v'$
could be different, which leads to somewhat different particle
physics in the mirror sector. The models with spontaneoulsy  
broken parity and their implications were considered in 
refs. \cite{BM,BDM,BV}. In this paper we mostly concentrate 
on the case with exact mirror parity.}
If the two sectors are separate and do not interact by forces 
other than gravity, the difference between $D$ and $P$ parities 
is rather symbolic and does not have any profound implications. 
However, in scenarios with some particle messengers 
between the two sectors this difference can be important 
and can have dynamical consequences.

\subsection{Couplings between O- and M-particles} 

Now we discuss, what common forces could exist between 
the O- and M-particles, including matter fields and 
gauge fields.  

$\bullet$ 
{\it Kinetic mixing term between the O- and M-photons} 
\cite{Holdom,Glashow86,Glashow87}. 
In the context of $G_{\rm SM} \times G_{\rm SM}'$, the 
general Lagrangian can contain the gauge invariant term 
%is no symmetry reason for suppressing the kinetic mixing term 
%between the field-strength tensors of the gauge
%factors $U(1)$ and $U(1)'$, 
\be{BBpr}
\cL = -\chi B^{\mu\nu} B'_{\mu\nu}, 
\ee
where $B_{\mu\nu} = \partial_\mu B_\nu - \partial_\nu B_\mu$,   
and analogously for $B'_{\mu\nu}$, where $B_\mu$ and $B'_\mu$ 
are gauge fields of the abelian gauge factors $U(1)$ and $U(1)'$. 
Obviously, after the electroweak symmetry breaking, this term 
gives rise to a kinetic mixing term between the field-strength 
tensors of the O- and M-photons: 
\be{FFpr} 
\cL = -\eps F^{\mu\nu} F'_{\mu\nu}
\ee
with $\eps = \xi \cos^2\theta_W$. There is no symmetry reason 
for suppressing this term, and generally the constant $\eps$ 
could be of order 1. 

On the other hand, experimental limits on 
the orthopositronium annihilation imply  
a strong upper bound on $\eps$.  
This is because one has to diagonalize first the kinetic 
terms of the $A_\mu$ and $A'_\mu$ states and identify the physical 
photon as a certain linear combination of the latter. 
One has to notice that after the kinetic terms are 
brought to canonical form by diagonalization and scaling 
of the fields, $(A,A')\to (A_1,A_2)$,  
any orthonormal combination of states $A_1$ and $A_2$ 
becomes good to describe the physical basis.  
In particular, $A_2$ can be chosen as a "sterile" state 
which does not couple to O-particles but only to M-particles. 
Then, the orthogonal combination $A_1$ couples not only 
to O-particles, but also with M-particles with a small 
charge $\propto 2\eps$ -- in other words, mirror matter 
becomes "milicharged" with respect to the physical ordinary 
photon \cite{Holdom,FYV}.  
In this way, the term (\ref{FFpr}) should induce the process
$e^+e^-\to e^{\prime+}e^{\prime-}$, with an amplitude just 
$2\eps$ times the $s$-channel 
amplitude for $e^+e^-\to e^+e^-$. 
By this diagram, orthopositronium would oscillate   
into its mirror counterpart, which would be seen as 
an invisible decay mode exceeding  
experimental limits  
unless $\eps < 5\times 10^{-7}$ or so \cite{Glashow86}.  

For explaining naturally the smallness of the kinetic mixing term
(\ref{FFpr}) one needs to invoke the concept of grand unification. 
Obviously, the term (\ref{BBpr}) is forbidden
if $G_{SM}\times G'_{SM}$ is embedded 
in GUTs like $SU(5)\times SU(5)'$ or $SO(10)\times SO(10)'$ 
which do not contain abelian factors. 
However, given that both $SU(5)$ and $SU(5)'$ symmetries are 
broken down to their $SU(3)\times SU(2)\times U(1)$ subgroups 
by the Higgs 24-plets $\Phi$ and $\Phi'$, it could emerge from 
the higher order effective operator 
\be{GGpr}
\cL= -\frac{\zeta}{M^2} (G^{\mu\nu} \Phi) (G'_{\mu\nu} \Phi') 
\ee
where $G_{\mu\nu}$ and $G'_{\mu\nu}$ are field-strength 
tensors respectively of $SU(5)$ and $SU(5)'$, and $M$ is 
some cutoff scale which can be of the order of $M_{Pl}$ or so. 
After substituting VEVs of $\Phi$ and $\Phi'$ 
%which are order GUT scale  
the operator (\ref{FFpr}) is induced  with 
$\eps \sim \zeta (\langle\Phi\rangle/M)^2$. 

In fact, the operator (\ref{GGpr}) can be effectively induced 
by loop-effects involving some heavy fermion or scalar 
fields in the mixed representations of $SU(5)$ and $SU(5)'$, 
with $\zeta \sim \al/3\pi$ being a loop-factor.   
Therefore, taking for the GUT scale 
$\langle\Phi\rangle\sim 10^{16}$ GeV and $M\sim M_{Pl}$ 
we see that if the kinetic mixing term (\ref{FFpr}) is 
induced at all, its natural range can vary from 
$\eps \sim 10^{-10}$ up to the upper limit of $5\times 10^{-7}$. 

For $\eps\sim {\rm few}\times 10^{-7}$ the term has  
striking experimental implications for positronium physics. 
Namely, the $e^+e^-\to e^{\prime+}e^{\prime-}$ 
process, would have an amplitude just $2\eps$ times the $s$-channel 
one for $e^+e^-\to e^+e^-$, and this would lead to mixing 
of ordinary positronium to its mirror counterpart 
with significant rate, and 
perhaps could help in solving the troubling mismatch 
problems in the positronium physics 
\cite{Glashow86,Gninenko94,Gninenko00}. 
However, this value is an order of magnitude above 
the limit $\eps < 3\times 10^{-8}$ from the BBN 
constraints\cite{Glashow87}. 
For larger $\eps$ the reaction $e^+e^-\to
e^{\prime+}e^{\prime-}$  
would funnel too much energy density from the ordinary to 
the mirror world and would violate the BBN limit on $\DN$.  
%Somewhat stronger bounds $\eps < 10^{-9.5}$ 
%can be obtained from the supernova core cooling rate 
%due to mirror photon production by the reaction 
%$\ga + e \to e + \ga'$ \cite{KMT96}. 

The search of the process $e^+e^- \to$ {\it invisible} 
could approach sensitivities down to few $\times 10^{-9}$. 
\cite{Sergei} This interesting experiment could test 
the proposal of ref. \cite{Feet} claiming that the signal 
for the dark matter detection by the DAMA/NaI group \cite{Rita} 
can be explained by elastic scattering of M-baryons with 
ordinary ones mediated by kinetic mixing (\ref{FFpr}), 
if $\eps \sim 4\times 10^{-9}$.

$\bullet$ 
{\it Mixing term between the O- and M-neutrinos}. 
In the presence of the M-sector, 
the $D=5$ operator responsible for neutrino masses 
(\ref{op-nu}) can be immediately generalized to include  
an analogous terms for M-neutrinos as well 
as the mixed terms between the O- and M-neutrinos. 
\be{op-gen}
\frac{A_{ij}}{2 M} (l_i \phi) (l_j \phi) +
\frac{\Apr_{ij}}{2 M} (\lpr_i\phpr) (\lpr_j \phpr)
+ \frac{D_{ij}}{M} (l_i\phi) (\lpr_j \phpr) 
+ {\rm h.c.} \;,
\ee
The first operator in eq.\ (\ref{op-gen}), due to the 
ordinary Higgs vacuum VEV $\langle\phi \rangle = v\sim 100$ GeV,
then induces the small Majorana masses of the ordinary
(active) neutrinos. 
Since the mirror Higgs $\phi^\prime$ also has a
non-zero VEV $\langle\phi^\prime \rangle = v^\prime$,
the second operator then provides the masses
of the M-neutrinos 
(which in fact are sterile for the ordinary observer),
and finally, the third operator induces the mixing mass terms
between the active and sterile neutrinos.  
The total mass matrix of neutrinos $\nu\subset l$
and $\nu^\prime\subset l^\prime$ reads as~\cite{BM}:
\be{numass}
M_\nu = \mat{m_\nu}{m_{\nu\nu^\prime}}{m_{\nu\nu^\prime}^T}
{m_{\nu^\prime}} =
\frac{1}{M} \mat{Av^2}{Dv v^\prime}{D^Tv v^\prime}
{A^\prime v^{\prime 2}} \, .
\ee
Thus, this model provides a simple explanation of
why sterile neutrinos could be light
(on the same grounds as the active neutrinos)
and could have significant mixing with the ordinary neutrinos.

The heavy neutrinos of the O-and M-sectors, 
$\tN$ and $\tN'$, are gauge singlets, and there is 
no essential difference between them. Therefore 
one can join states like $\tN$ and $\tN'$ in a generalized  
set of gauge singlet fermions $N_a$ ($a=1,2,..$). 
They all could mix with each other, and, 
in the spirit of the seesaw mechanism,
they can couple to leptons of both O- and M-sectors. 
The relevant Yukawa couplings have the form:
\be{Yuk-nu}
Y_{ia}l_i N_a \phi + Y^\prime_{ia}\lpr_i N_a \phpr +
\frac{1}{2} M_{ab} N_a N_b
+  {\rm h.c.}
\ee
%(charge-conjugation matrix $C$ is omitted);
In this way, $N$ play the role of messengers between ordinary
and mirror particles.
After integrating out these heavy states, the operators 
(\ref{op-gen}) are induced with $A=YG^{-1}Y^T$, 
$A'=Y'G^{-1}Y^{\prime T}$ and $D=YG^{-1}Y^{\prime T}$. 
In the next section we show that in addition the $N$ 
states can mediate L and CP violating scattering processes 
between the O- and M-sectors which could provide a new  
mechanism for primordial leptogenesis. 
 
It is convenient to present the heavy neutrino mass matrix
as $M_{ab} = G_{ab} M$,
$M$ being the overall mass scale and $G_{ab}$  
some typical Yukawa constants.
Without loss of generality, $G_{ab}$
can be taken diagonal and real. Under $P$ or $D$ parities,  
in general some of the states 
$N_a$ would have positive parity, while others could have 
a negative one. 

One the other hand, the Yukawa matrices in general remain 
non-diagonal and complex. Then $D$ parity would 
imply that $Y'=Y$, while $P$ parity imples $Y'=Y^\ast$ 
(c.f. (\ref{DLL1}) and (\ref{PLR1})).  

%All fermion states $l,N,l^\prime$ are taken to be left-handed
%while their $C$-conjugate,
%right-handed anti-particles are denoted as
%$\bar{l},\bar{N},\bar{l}^\prime$.   

$\bullet$ 
{\it Interaction term between the O- and M- Higgses }. 
In the context of $G_{\rm SM} \times G_{\rm SM}'$, the gauge 
symmetry allows also a quartic interaction term between the 
O-and M-Higgs doublets $\phi$ and $\phi'$: 
\be{HHpr} 
\la (\phi^\dagger \phi)(\phi^{\prime\dagger}\phi') 
\ee 
This term is cosmologically dangerous, since it would bring 
the two sectors into equilibrium in the early Universe via 
interactions $\bar{\phi}\phi \to \bar{\phi}'\phi'$ 
unless $\la$ is unnaturally small, $\la < 10^{-8}$.\cite{BDM}

However, this term can be properly suppressed by supersymmetry. 
In this case standard Higgses $\phi_{u,d}$ become chiral 
superfields as well as their mirror partners $\phi'_{u,d}$, 
and so the minimal gauge invariant term between the 
O- and M-Higgses in the superpotential has dimension 5: 
$(1/M)(\phi_u\phi_d) (\phi'_u\phi'_d)$, where $M$ is some 
big cutoff mass, e.g. of the order of the GUT scale or Planck scale.    
Therefore, the general Higgs Lagrangian takes the form:
\be{susyHH} 
{\cal L} = \int d^2\theta 
[\mu \phi_u\phi_d + \mu \phi'_u\phi'_d
+ \frac{1}{M} (\phi_u\phi_d) (\phi'_u\phi'_d)] ~ + ~ {\rm h.c.}
\ee
plus unmixed D-terms, where $\mu$-terms are of the order of 100 GeV 
and $M$ is some large cutoff scale. 
This Lagrangian contains 
a mixed quartic terms similar to (\ref{HHpr}): 
\be{HH}
\la (\phi_u^\dagger\phi_u)(\phi'_u\phi'_d) + 
\la (\phi_d^\dagger\phi_d)(\phi'_u\phi'_d) + 
(\phi_{u,d}\to \phi'_{u,d}) + {\rm h.c.} 
\ee
with the coupling constant $\la = \mu/M$. The same holds true 
for the soft supersymmetry breaking $F$-term and $D$-terms. 
For example, the F-term 
$\frac{1}{M}\int d^2\theta z (\phi_u \phi_d)(\phi'_u \phi'_d) + $ h.c.,  
where $z=m_S\theta^2$ being the supersymmetry breaking spurion
with $m_S \sim 100$ GeV, gives rise to a quartic scalar term 
\be{HH1}
\la (\phi_u\phi_d)(\phi'_u\phi'_d) + {\rm h.c.} 
\ee
with $\la \sim m_S/M \ll 1$. Thus for $\mu,m_S\sim 100$ GeV, 
all these quartic constants are strongly suppressed, and 
hence are safe. 

$\bullet$ {\it Mixed multiplets between the two sectors.} 
Until now we discussed the situation with O-particles 
being singlets of mirror gauge factor $G'$ and vice versa, 
M-particles being singlets of ordinary gauge group $G$. 
However, in principle there could be also some fields 
in mixed representations of $G\times G'$. Such fields 
usually emerge if the two gauge factors $G$ and $G'$ 
are embedded into a bigger grand unification group $\cG$. 

For example, consider a gauge theory $SU(5)\times SU(5)'$ 
where the O- and M-fermions respectively are in the 
following left-chiral multiplets: 
$L \sim (\bar5+10,1)$ and $L'\sim (1,5+\ov{10})$.  
One can introduce however also the left chiral fermions 
in mixed representations like $F\sim (5,5)$ and 
$F'\sim (\bar5,\bar5)$, $T\sim (\ov{10},\ov{10})$ and 
$T'\sim (10,10)$, etc. 
Mixed multiplets would necessarily appear 
if $SU(5)\times SU(5)'$ is embbedded into $SU(10)$ group. 
They should have a large mass term $M$, e.g. of the order of 
$SU(10)$ breaking scale to $SU(5)\times SU(5)'$. 
However, in general they could couple also to the GUT Higgses 
$\Phi\sim (24,1)$ and $\Phi'\sim (1,24)$. Thus, their Lagrangian 
takes a form  
$MFF' + \Phi FF' + \Phi'FF' +$ h.c. 
%$MFF' + M\ast \bar{F}\bar{F}'$ etc. 
%Mirror parity $P$ (\ref{PLR}) requires that mass terms $M$ are
%real (or, in case of many families, they are hermitean matrices). 
%fields should transform as 
%$f,t \to \bar{f}',\bar{t}'$, $F,T\to $

%$\langle\Phi\rangle = V{\rm diag}(-1/3,-1/3,-1/3,1/2,1/2)$ 

Now, under the $G_{\rm SM}\times G_{\rm SM}'$ subgroup, 
these multiplets split into fragments $F_{ij}$ with different 
hypercharges $(Y_i,Y_j')$ and masses 
$\cM_{ij} = M+Y_i\langle\Phi\rangle + Y_j'\langle\Phi'\rangle$. 
Therefore, the loops involving 
the fermions $F_{ij}$ would induce a contribution to 
the term (\ref{BBpr}) which reads as 
$\chi = (\al/3\pi) {\rm Tr}[YY'\ln(\cM/\La)]$ 
where $\La$ is an ultraviolet cutoff scale and under trace 
the sum over all fragments $F_{ij}$ is understood. 
However, as far as these fragments emerge from the GUT 
multiplets, they necessarily obey that ${\rm Tr}(YY')=0$, 
and thus $\chi$ should be finite and cutoff independent. 
Thus, expanding the logarithm in terms of small parameters 
$\langle\Phi^{(\prime)}\rangle/M$, we finally obtain 
\be{BB-rad}
\chi = \frac{\al\langle\Phi\rangle \langle\Phi'\rangle}
{3\pi M^2} {\rm Tr}[(YY')^2]   
\ee 
exactly what we expected from the effective operator 
(\ref{GGpr}). Hence, the heavy mixed multiplets in fact 
do not decouple and induce the O- and M-photon kinetic 
mixing term proportional to the square of typical mass 
splittings in these multiplets ($\sim \langle\Phi\rangle^2$), 
analogously to the familiar situation for the photon to 
$Z$-boson mixing in the standard model.

$\bullet$ {\it Interactions via common gauge bosons.} 
It is pretty possible that O-and M-particles have 
common forces mediated by the gauge bosons of some 
additional symmetry group $H$. 
In other words,  
one can consider a theory with a gauge group 
$G\times G'\times H$, where O-particles are 
in some representations of $H$, $L_a\sim r_a$, and 
correspondingly their antiparticles are in antirepresentations, 
$\tR_a \sim \bar{r}_a$. 
As for M-particles, vice versa, we take $L'_a\sim \bar{r}_a$, 
and so $\tR'_a \sim r_a$. Only such a prescription of $\cG$ 
pattern is compatible with the mirror parity (\ref{PLR}). 
In addition, in this 
case $H$ symmetry automatically becomes vector-like and 
so it would have no problems with axial anomalies even 
if the particle contents of O- and M-sectors separately are 
not anomaly-free with respect to $H$. 

Let us consider the following example. The horizontal 
flavour symmetry $SU(3)_H$ between the quark-lepton families 
seems to be very promising for understanding the fermion 
mass and mixing pattern \cite{su3,PLB83}. 
In addition, it can be   
useful for controlling the flavour-changing phenomena 
in the context of supersymmetry \cite{PLB98}.      
One can consider e.g. a GUT with $SU(5)\times SU(3)_H$ 
symmetry where L-fermions in (\ref{LL-SM}) are triplets of
$SU(3)_H$. So $SU(3)_H$ has a chiral character and 
it is not anomaly-free unless some extra states are introduced 
for the anomaly cancellation \cite{su3}.  

However, the concept of mirror sector makes the things 
easier. 
Consider e.g. $SU(5)\times SU(5)'\times SU(3)_H$ theory  
with $L$-fermions in (\ref{LL-SM}) being triplets of $SU(3)_H$ 
while $L'$-fermions in (\ref{LL-SMpr}) are anti-triplets. 
Hence, in this case the $SU(3)_H$ anomalies of the ordinary  
particles are cancelled by their mirror partners.  
Another advantage is that in a supersymmetric theory 
the gauge D-terms of $SU(3)_H$ are perfectly cancelled between 
the two sectors and hence they do not give rise to dangerous 
flavour-changing phenomena \cite{PLB98}.  

The immediate implication of such a theory would be the 
mixing of neutral O-bosons to their M-partners, mediated 
by horizontal gauge bosons. Namely, oscillations 
$\pi^0 \to \pi^{\prime 0}$ or $K^0 \to K^{\prime 0}$ 
become possible and perhaps even detectable if the 
horizontal symmetry breaking scale is not too high. 

Another example is a common lepton number (or $B-L$) 
symmetry between the two sectors. Let us assume that ordinary 
leptons $l$ have lepton charges $L=1$ under this symmetry 
while mirror ones $l'$ have $L=-1$. Obviously, this symmetry 
would forbid the first two couplings in (\ref{op-gen}),       
$A,A'=0$, while the third term is allowed -- $D\neq 0$.  
Hence, `Majorana' mass terms would be absent both for 
O- and M-neutrinos in (\ref{numass}) and so  
neutrinos would be Dirac particles having 
{\it naturally small} masses, with left components  
$\nu_L\subset l$ and right components being 
$\tnu'_R\subset \tl$.    

The model with common Peccei-Quinn symmetry between the 
O- and M-sectors was considered in \cite{BGG}.

\section{The expansion of the Universe and thermodynamics of the O- and
M-sectors}

Let us assume, that after inflation ended, the 
O- and M-systems received different reheating temperatures,  
namely $T_R> T'_R$. This is certainly possible despite 
the fact that two sectors have identical Lagrangians, and 
can be naturally achieved in certain 
models of inflation \cite{KST,BDM,BV}.\footnote{For analogy, 
two harmonic oscillators with the same frequency 
(e.g. two springs with the same material and the same length) 
are not obliged to oscillate with the same amplitudes.} 
  
If the two systems were decoupled already
after reheating, at later times $t$ they will have different
temperatures $T(t)$ and $T'(t)$, and so 
different energy and entropy densities:
\be{rho}
\rho(t) = {\pi^{2}\over 30} g_\ast(T) T^{4}, ~~~
\rho'(t) = {\pi^{2}\over 30} g'_\ast(T') T'^{4} ~,
\ee
\be{s}
s(t) = {2\pi^{2}\over 45} g_{s}(T) T^{3} , ~~~
s'(t) = {2\pi^{2}\over 45} g'_{s}(T') T'^{3} ~.
\ee
The factors $g_{\ast}$, $g_{s}$ and $g'_{\ast}$, $g'_{s}$
accounting for the effective number of the degrees of freedom
in the two systems can in general be different from each other.
Let us assume that during the expansion of the Universe 
the two sectors evolve with separately conserved entropies. 
Then the ratio $x\equiv (s'/s)^{1/3}$ is time independent 
while the ratio of the temperatures
in the two sectors is simply given by:
\be{t-ratio}
\frac{T'(t)}{T(t)} = x \cdot
\left[\frac{g_{s}(T)}{g'_{s}(T')} \right] ^{1/3} ~.
\ee

The Hubble expansion rate is determined by the total
energy density $\rhb=\rho+\rho'$, $H=\sqrt{(8\pi/3) G_N\rhb}$.
Therefore, at a given time $t$ in a radiation dominated epoch
we have
\be{Hubble}
H(t) = {1\over 2t} = 1.66 \sqrt{\bg(T)} \frac{T^2}{M_{Pl}} =
1.66 \sqrt{\bg'(T')} \frac{T'^2}{M_{Pl}} ~
\ee
in terms of O- and M-temperatures $T(t)$ and $T'(t)$, where
\beqn{g-ast}
\bg(T) = g_\ast (T) (1 + x^4), ~~~~
\bg'(T') = g'_\ast (T')(1 + x^{-4} ) .
\eeqn
%Here the factor $a(T,T') = [g'_\ast (T')/g_\ast (T)]
%\cdot [g_{s}(T)/g'_{s}(T')]^{4/3}$
%takes into account that for $T'\neq T$ the relativistic
%particle contents of the two worlds can be different.
%However, except for very small values of $x$,
%we have $a \sim 1$. So hereafter we always take
%$\bg(T) = g_\ast (T) (1 + x^4)$ and
%$\bg'(T') = g'_\ast (T')(1 + x^{-4})$.

In particular, we have  $x= T'_0/T_0$,
where $T_0,T'_0$ are the present
temperatures of the O- and M- relic photons.
In fact, $x$ is the only free parameter in our model
and it is constrained by the BBN bounds.

The observed abundances of light elements are in
good agreement with the standard nucleosynthesis predictions.
At $T\sim 1$ MeV we have $g_\ast=10.75$
as it is saturated by photons $\gamma$, electrons $e$  
and three neutrino species $\nu_{e,\mu,\tau}$.
The contribution of mirror particles
($\gamma'$, $e'$ and $\nu'_{e,\mu,\tau}$)
would change it to $\bg =g_\ast (1 + x^4)$.
Deviations from $g_\ast=10.75$ are usually
parametrized in terms of the effective number
of extra neutrino species,
$\Delta g= \bar{g}_\ast -10.75=1.75\Delta N_\nu$.
Thus we have:
\be{BBN}
\Delta N_\nu = 6.14\cdot x^4 ~.
\ee
This limit very weakly depends on $\DN$.
Namely, the conservative bound $\Delta N_\nu < 1$ 
implies $x < 0.64$. 
In view of the present observational situation,
confronting the WMAP results to the BBN analysis, 
the bound seems to be stronger. 
However, e.g. $x = 0.3$ implies a completely negligible 
contribution  $\Delta N_\nu = 0.05$. 

As far as $x^4\ll 1$, in a relativistic epoch
the Hubble expansion rate (\ref{Hubble}) is dominated
by the O-matter density and the presence of the M-sector
practically does not affect the standard cosmology
of the early ordinary Universe.
However, even if the two sectors have the
same microphysics, the cosmology of the early
mirror world can be very different from the
standard one as far as the crucial epochs like 
baryogenesis, nuclesosynthesis, etc. are concerned.
Any of these epochs is related to an instant when
the rate of the relevant particle process
$\Ga(T)$, which is generically a function
of the temperature, becomes equal to the Hubble
expansion rate $H(T)$.
Obviously, in the M-sector these events take place
earlier than in the O-sector,
%since the former is colder than the later,
and as a rule, the relevant processes in the former
freeze out at larger temperatures than in the latter.

In the matter domination epoch the situation becomes
different.
%At present it is under discussion to which is 
%the non-relativistic matter fraction $\Omega_m$   
%in the universe.
In particular, we know that ordinary baryons provide
only a small fraction of the present matter density,
%$\Omega_B \simeq 0.04$, 
whereas the observational data
indicate the presence of dark matter with about 5 times 
larger density.  
%can amount for $\Omega_{m} \simeq 0.25$.
It is interesting to question whether the missing
matter density of the Universe could be due to
mirror baryons? In the next section we show that   
this could occur in a pretty natural manner.

It can also be shown that the BBN epoch in the mirror
world proceeds differently from the ordinary one,
and it predicts different abundancies of primordial 
elements \cite{BCV}. 
Namely, mirror helium abundance can be in the
range $Y'_4 =0.6-0.8$, considerably larger than
the observable $Y_4\simeq 0.24$.

\section{Baryogenesis in M-sector and mirror baryons as 
dark matter} 

\subsection{Visible and dark matter in the Universe}

The present cosmological observations strongly support 
the main predictions of the inflationary scenario: 
first, the Universe is flat, with the energy density very close 
to the critical $\Om=1$, and second, 
primoridal density perturbations have nearly flat spectrum, 
with the spectral index $n_s\approx 1$. 
The non-relativistic matter gives only a small 
fraction of the present energy density, about 
$\Om_m \simeq 0.27$, while the rest is attributed to the 
vacuum energy (cosmological term) $\Om_\La \simeq 0.73$ 
\cite{WMAP}. 
The fact that $\Om_m$ and $\Om_\La$ are of the same order, 
gives rise to so called cosmological coincidence problem: 
why we live in an epoch when $\rho_m\sim\rho_\la$, if in the 
early Universe one had $\rho_m \gg \rho_\La$ and in the 
late Universe one would expect $\rho_m\ll \rho_\La$? 
The answer can be only related to an antrophic principle: 
the matter and vacuum energy densities 
scale differently with the expansion of the Universe 
$\rho_m\propto a^{-3}$ and $\rho_\La \propto$ const., 
hence they have to coincide at some moment, 
and we are just happy to be here. Moreover, for 
substantially larger $\rho_\La$ no galaxies could be formed 
and thus there would not be anyone to ask this this question. 

On the other hand, the matter in the Universe has two 
components, visible and dark: $\Om_m = \Om_b +\Om_{d}$.  
The visible matter consists of baryons  
with $\Om_b \simeq 0.044$ while the dark matter 
with $\Om_{d} \simeq 0.22$ is constituted by some hypothetic 
particle species very weakly interacting with the 
observable matter. 
It is a tantalizing question, why the visible and dark 
components have so close energy densities?  
Clearly, the ratio  
\be{b-dm}
\beta = \frac{\rho_d}{\rho_b}   
\ee 
does not depend on time as far as with the expansion of the 
Universe both $\rho_b$ and $\rho_{d}$ scale as $\propto a^{-3}$. 

%$\Om_b/\Om_{dm}\simeq 0.2$

In view of the standard cosmological paradigm,
there is no good reason for having $\Om_d \sim \Om_b$, 
as far as the visible and dark components have  
different origins. 
The density of the visible matter is $\rho_b =M_N n_b$,     
where $M_N\simeq 1$ GeV is the nucleon mass, and 
$n_b$ is the baryon number density of the Universe.   
The latter should be produced in a very early Universe 
by some baryogenesis mechanism, which is presumably related 
to some B and CP-violating physics at very high energies. 
The baryon number per photon $\eta=n_b/n_\ga$ is very small. 
Observational data on the primordial abundances of light 
elements and the WMAP results on the CMBR anisotropies nicely 
converge to the value $\eta \approx 6\times 10^{-10}$.  

As for dark matter, it is presumably constituted by 
some cold relics with mass $M$ and number density $n_{d}$, 
and $\rho_{d} = M n_{d}$. 
The most popular candidate for cold dark matter (CDM) 
is provided by the lightest supersymmetric particle (LSP) 
with $M_{\rm LSP}\sim 1$ TeV, 
and its number density $n_{\rm LSP}$ 
is fixed by its annihilation cross-section. 
%which strongly depends on the model parameters,  
Hence $\rho_b\sim \rho_{\rm LSP}$ requires that   
$n_b/n_{\rm LSP}\sim M_{\rm LSP}/M_N$ and the 
origin of such a conspiracy between four principally independent 
parameters is absolutely unclear.  
Once again, the value $M_N$ is fixed by the QCD scale while 
$M_{\rm LSP}$ is related to the supersymmetry breaking scale, 
$n_b$ is determined by B and CP violating properties of 
the particle theory at very high energies whereas  
$n_{\rm LSP}$ strongly depends on the supersymmetry 
breaking details. Within the parameter space of the MSSM 
it could vary within several orders of magnitude, and moreover, 
in either case it has nothing to do with the B and CP 
violating effects.  

The situation looks even more obscure if the dark component 
is related e.g. to the primordial oscillations 
of a classic axion field, in which case the dark matter 
particles constituted by axions are superlight, with mass 
$\ll 1$ eV, but they have a condensate with enormously 
high number density. 

In this view, the concept of mirror world could give 
a new twist to this problem. Once the visible matter 
is built up by ordinary baryons, then the mirror 
baryons could constitute dark matter in a natural way.  
They interact with mirror photons, however they are dark 
in terms of the ordinary photons. 
The mass of M-baryons is the same as the ordinary 
one, $M=M_N$, and so we have $\beta = n'_b/n_b$, 
where $n'_b$ is the number density of M-baryons. 
In addition, as far as the two sectors have the same 
particle physics, it is natural to think that the M-baryon 
number density $n'_b$ is determined by the baryogenesis 
mechanism which is similar to the one which fixes the 
O-baryon density $n_b$. Thus, one could question whether 
the ratio $\beta=n'_b/n_b$ could be naturally order 1 
or somewhat bigger.     

The visible matter in the Universe consists of baryons, 
while the abundance of antibaryons is vanishingly small. 
In the early Universe, at tempreatures $T\gg 1$ GeV, 
the baryons and antibaryons had practically the same 
densities, $n_b\approx n_{\bar b}$ with $n_b$ slightly 
exceeding $n_{\bar b}$, so that 
the baryon number density was small, 
$n_B=n_b-n_{\bar b}\ll n_b$. 
If there was no significant entropy production after 
the baryogenesis epoch,  
the baryon number density to entropy density 
ratio had to be the same as today,  
$B = n_B/s \approx 8\times 10^{-11}$.\footnote{
In the following we use $B=n_B/s$ which is related  
with the familiar $\eta=n_B/n_\ga$ as $B\approx 0.14\eta$. 
However, $B$ is more adopted for featuring the baryon 
asymmetry since it does not depend on time if the entropy of 
the Universe is conserved.}
 
One can question, who and how has prepared the initial 
Universe with such a small excess of baryons over antibaryons. 
In the Friedman Universe the initial baryon asymmetry 
%excess of baryons over antibaryons 
could be arranged a priori, in terms of non-vanishing 
chemical potential of baryons. 
However, the inflationary paradigm gives another 
twist to this question, since inflation dilutes any 
preexisting baryon number of the Universe to zero.  
Therefore, after inflaton decay and the (re-)heating of the Universe, 
the baryon asymmetry has to be created by some cosmological 
mechanism.  

There are several relatively honest baryogenesis mechanisms 
as are GUT baryogenesis, leptogenesis, electroweak baryogenesis, 
%Affleck-Dine mechanism 
etc. (for a review, see e.g. \cite{BA-Dolgov}). 
They are all based on general principles 
suggested long time ago by Sakharov \cite{Sakh}:  
a non-zero baryon asymmetry can be produced in the initially baryon
symmetric Universe if three conditions are fulfilled: B-violation,
C- and CP-violation and departure from thermal equilibrium.
In the GUT baryogenesis or leptogenesis scenarios 
these conditions can be satisfied in the decays of heavy 
particles. 
%of grand unified theories. 

At present it is not possible to say definitely which 
of the known mechanisms is responsible 
for the observed baryon asymmetry in the ordinary world. 
However, it is most likely that the baryon asymmetry in 
the mirror world 
%$\eta'=n'_B/n'_\gamma$ 
is produced by the same mechanism
and moreover, the properties of the $B$ and CP violation
processes are parametrically the same in both cases.
But the mirror sector has a lower temperature than ordinary one, 
and so at epochs relevant for baryogenesis the out-of-equilibrium 
conditions should be easier fulfilled for the M-sector.

%Present cosmological observations show that the baryon number 
%per photon is 
%The measurement of the primordial abundances of light elements 
%and the WMAP results on the CMBR anisotropies nicely converge 
%to the value $\eta = n_B/n_\ga \approx 6\times 10^{-10}$.  
%It is more convenient to express baryon asymmetry of the 
%Universe in terms of the baryon density to the entropy density, 
%$B=n_B/s$, given the fact that this ratio does not depend on 
%time during the late stages of the evolution of the early universe, 
%i.e. it is the same at BBN epoch and in the present universe.
%t present $s=3.91n_\ga$, and so we have 
%$B=0.14 \eta = 8.4 \times 10^{-11}$. 

\subsection{Baryogenesis in the O- and M-worlds} 

Let us consider the difference between the ordinary and 
mirror baryon asymmetries on the example of the 
GUT baryogenesis mechanism. It is typically based on
`slow' B- and CP-violating decays of a superheavy boson $X$
into quarks and leptons, where slow means that    
at $T < M$ the Hubble parameter $H(T)$ is greater than   
the decay rate $\Gamma \sim \alpha M$, 
$\alpha$ being the coupling strength of $X$ to fermions
and $M$ its mass. 
The other reaction rates are also of relevance: 
{\it inverse decay}: $\Gamma_{I} \sim  \Gamma    
(M/T)^{3/2} \exp(-M/T)$ for $T< M_X$, 
and {\it the $X$ boson mediated scattering processes}:
$\Gamma_S \sim n_X \sigma\sim  A \alpha^2T^5/M^4$,
where the factor $A$ amounts for the possible reaction channels.

The final BA depends on the temperature
at which $X$ bosons go out from equilibrium.
%and on which among the above reactions is
%most effective in damping baryon asymmetry.
One can introduce a parameter which measures the
effectiveness of the decay at the
epoch $T\sim M$:
$k=(\Gamma/H)_{T=M}= 0.3\bg^{-1/2}(\alpha M_{Pl}/M)$.
For $k \ll 1$ the out-of-equilibrium condition is 
strongly satisfied, and per decay of one $X$ particle 
one generates the baryon number proportional to the 
CP-violating asymmetry $\eps$. Thus,  
we have $B = \eps/g_\ast$, 
%where $\eps$ is a CP-violating asymmetry in $X$-boson decay and 
$g_\ast$ is a number of effective degrees of freedom at $T < M$.   
The larger $k$ is, the longer equilibrium is
maintained and the freeze-out abundance of $X$ boson
becomes smaller. 
Hence, the resulting
baryon number to entropy ratio becomes  
$B=(\eps/g_\ast)D(k)$, where the damping factor $D(k)$  
is a decreasing function of $k$. 
%$B=n_B/s\simeq 0.14\eta$, 
In particular, $D(k)=1$ for $k\ll 1$, while 
for $k$ exceeding some critical value 
$k_c$, the damping is exponential.  

The presence of the mirror sector practically  
does not alter the ordinary baryogenesis.
The effective particle number is
$\bar g_\ast (T) = g_\ast(T)(1+x^4)$ and thus
the contribution of M-particles to the Hubble constant  
at $T\sim M$ is suppressed by a small factor $x^4$.

In the mirror sector everything should occur
in a similar way, apart from the fact that now  
at $T'\sim M$ the Hubble constant is not
dominated by the mirror species but by ordinary ones:
$\bar g'_\ast (T')\simeq g'_\ast (T')(1+ x^{-4})$.
As a consequence, we have
$k' = (\Gamma/H)_{|T'=M} = k x^2$.
%Since the value of $k_c$ is the same in the two sectors,
Therefore, the damping factor for mirror baryon asymmetry 
can be simply obtained by replacing 
$k\rightarrow k'=kx^2$ in $D(k)$. 
In other words, the baryon number density to entropy density 
ratio in the M-world becomes 
$B'=n'_B/s' \simeq (\epsilon/g_\ast) D(kx^2)$.
Since $D(k)$ is a decreasing function of $k$, then
for $x < 1$ we have $D(kx^2) > D(k)$ and
thus we conclude that the mirror world always gets a
{\it larger} baryon asymmetry than the visible one, 
$B' > B$.\footnote{As it was shown in ref. \cite{BCV}, 
the relation $B' > B$ takes place also in the context 
of the electroweak baryogenesis scenario, where the 
out-of-equilibrium conditions is provided by fast 
phase transition and bubble nucleation.} 
Namely, for $k>1$ the baryon asymmetry in the O-sector is 
damped by some factor -- we have 
$B\simeq (\eps/g_\ast)D(k) < \eps/g_\ast$, 
while if $x^2 < k^{-1}$, the damping would be irrelevant 
for the M-sector and hence $B'\simeq \eps/g_\ast$.  

However, this does not a priori mean
that $\Omega'_b$ will be larger than $\Omega_b$.
Since the entropy densities are related as $s'/s=x^3$,
for the ratio $\beta =\Omega'_b/\Omega_b$ we have:
\be{B-ratio}
\beta(x) =  \frac{n'_B}{n_B} = \frac{B's'}{Bs} =
\;\frac{x^3D(kx^2)}{D(k)} ~.
\ee
The behaviour of this ratio 
%factor $x^3 F(kx^2,k_c)$   
as a function of $k$ for different values of the
parameter $x$ is given in the ref. \cite{BCV}.
Clearly, in order to have $\Omega'_b > \Omega_b$, 
the function $D(k)$ has to decrease
faster than $k^{-3/2}$ between $k'=kx^2 $ and $k$.
Closer inspection of this function reveals
that the M-baryons can be overproduced only if
$k$ is sufficiently large, so that 
the relevant interactions in the observable sector
maintain equilibrium longer than in the mirror one,
and thus ordinary BA can be suppressed by an
exponential Boltzmann factor while the mirror BA  
could be produced still in the regime $k' =kx^2 \ll 1$, 
when $D(k')\approx 1$. 

However, the GUT baryogenesis picture has the 
following generic problem. 
In scenarios based on grand unification models like $SU(5)$, 
the heavy gauge or Higgs boson decays violate separately 
$B$ and $L$, but conserve $B-L$, and so finally $B-L=0$.
On the other hand, the non-perturbative sphaleron processes, 
which violate $B+L$ but conserve $B-L$,
are effective at temperatures from about $10^{12}$ GeV down to 
100 GeV \cite{KRS}. 
Therefore, if $B+L$ is erased by sphaleron transitions, 
the final $B$ and $L$ both will vanish. 

Hence, in a realistic scenario one actually has to produce 
a non-zero $B-L$ rather than just a non-zero $B$, 
a fact that strongly favours the so called {\sl leptogenesis} 
scenario \cite{FY}.
The seesaw mechanism for neutrino masses offers an elegant   
possibility of generating non-zero $B-L$ in CP-violating decays  
of heavy Majorana neutrinos $N$ into leptons and Higgses. 
These decays violate $L$ but obviously do not change $B$  
and so they could create a non-zero $B-L = - L_{\rm in}$.   
Namely, due to complex Yukawa constants, the decay rates
$\Gamma(N\to l\phi)$ and $\Gamma(N \to \tl\tphi)$      
can be different  from each other, so that the leptons $l$ and
anti-leptons $\tl$ are produced in different amounts.

When sphalerons are in equilibrium, they violate $B+L$ and 
so redistribute non-zero $B-L$ between the 
baryon and lepton numbers of the Universe. Namely, the final values  
of $B$ and $B-L$ are related as $B = a(B-L)$, where 
$a$ is order 1 coefficient, namely $a\simeq 1/3$ in the 
SM and in its supersymmetric extension \cite{BA-Dolgov}. 
Hence, the observed baryon to entropy density ratio,
$ B \approx 8 \times 10^{-11} $,
needs to produce $ B-L \sim 2\times 10^{-10} $.

However, the comparative analysis presented above for the 
GUT baryogenesis in the O- and M-worlds, is essentially
true also for the leptogenesis scenario. 
The out-of-equilibrium decays of heavy $N$ neutrinos of 
the O-sector would produce a non-zero $B-L$ which being reprocessed 
by sphalerons would give an observable baryon asymmetry 
$B=a(B-L)$. On the other hand, the same decays of heavy 
$N'$ neutrinos of the M-sector would give non-zero $(B'-L')$  
and thus the mirror baryon asymmetry $B'=a(B'-L')$. 
In order to thermally produce heavy neutrinos in 
both O- and M-sectors, the lightest of them should 
have a mass smaller than the reheating temperature $T'_R$ 
in the M-sector, i.e. $M_N < T'_R ,T_R$. 
The situation $M_N > T'_R$ 
would prevent thermal production of $N'$ states, 
and so no $B'-L'$ would be generated in M-sector.  
However, one can 
consider also scenarios when both $N$ and $N'$ states 
are non-thermally produced in inflaton decays, but 
with different amounts. Then the reheating of both sectors 
as well as $B-L$ number generation can be related to the 
decays of the heavy neutrinos of both sectors and hence 
the situation $T'_R < T_R$ can be naturally accompanied by 
$B' > B$.

\subsection{Baryogenesis via Ordinary-Mirror particle exchange} 

An alternative mechanism of leptogenesis based on scattering
processes rather than on decay was suggested in ref. 
\cite{BB-PRL}.
The main idea consists in the following.
There exists some hidden (shadow) sector of new particles  
which are not in thermal equilibrium with the ordinary particle
world as far as the two systems interact very weakly,
e.g.~if they only communicate via gravity.
However, other messengers may well exist,   
namely, superheavy gauge singlets like right-handed
neutrinos which can mediate very weak effective interactions
between the ordinary and hidden leptons.
Then, a net $ B-L $ could emerge in the Universe as
a result of CP-violating effects in the unbalanced production
of hidden particles from ordinary particle collisions.     

Here we consider the case when the hidden sector is a mirror one.
As far as the leptogenesis is concerned, we concentrate only
on the lepton sector of both O and M worlds.
Therefore we consider the standard model, and among
other particles species, concentrate on 
the lepton doublets $l_i = (\nu, e)_i$
($ i=1,2,3 $ is the family index)
and the Higgs doublet $ \phi $ for the O-sector, 
and on their mirror partners 
$ l'_i = (\nu', e')_i $ and $ \phi' $. 
Their couplings to the heavy singlet neutrinos are 
given by (\ref{Yuk-nu}). 

Let us discuss now the leptogenesis mechanism in our scenario.
A crucial role in our considerations is played by the reheating
temperature $T_R$, at which the inflaton decay and entropy production
of the Universe is over, and after which the Universe is dominated by
a relativistic plasma of ordinary particle species.
As we discussed above, we assume that after the postinflationary
reheating, different temperatures are established in the two
sectors: $T'_R < T_R$,  i.e. the mirror sector is cooler than
the visible one, or ultimately, even completely ``empty".

In addition, the two particle systems should interact very weakly
so that they do not come in thermal equilibrium with each other
after reheating.
We assume that the heavy neutrino masses are larger than 
$T_R$ and thus cannot be thermally produced.
As a result, the usual leptogenesis mechanism via
$N\to l\phi$ decays is ineffective.

Now, the important role is played by lepton number violating
scatterings mediated by the heavy neutrinos $ N $.
The ``cooler" mirror world starts to be ``slowly" occupied
due to the entropy transfer from the ordinary sector through   
the $ \Delta L=1 $ reactions
$ l_i \phi \to \bar \lpr_k \barphpr $,
$ \bar l_i \barphi \to \lpr_k \phpr $.
In general these processes violate CP due to complex
Yukawa couplings in eq.~(\ref{Yuk-nu}), and so the
cross-sections with leptons and anti-leptons in the initial state
are different from each other.
As a result, leptons leak to the mirror sector more (or less)
effectively than antileptons and a non-zero $B-L$ is produced
in the Universe. 

It is important to stress that this mechanism would generate
the baryon asymmetry not only in the observable sector,
but also in the mirror sector. In fact, the two sectors are completely
similar, and have similar CP-violating properties.
We have scattering processes which transform the ordinary particles
into their mirror partners, and CP-violation effects in this
scattering owing to the complex coupling constants.
These exchange processes are active at some early epoch of the 
Universe, and they are out of equilibrium.
In this case, a hypothetical O observer should detect during the
contact epoch that
(i) matter slowly (in comparison to the Universe expansion rate)  
disappears from the thermal bath of our world, and, in addition,
(ii) particles and antiparticles disappear with different rates,
so that after the contact epoch ends up, he observes that his world
is left with a non-zero baryon number even if initially it was 
baryon symmetric.

On the other hand, his mirror colleague, M observer, would see that
(i) matter creation takes place in his world, and
(ii) particles and antiparticles emerge with different rates.  
Therefore, after the contact epoch, he also would observe
a non-zero baryon number in his world.

One would naively expect that in this case the baryon asymmetries
in the O and M sectors should be literally equal,
given that the CP-violating factors are the same for both sectors.
However, we show that in reality, the BA in the M sector,
since it is colder, can be about an order of magnitude bigger   
than in the O sector, as far as washing out effects are taken into account.
Indeed, this effects should be more efficient for the hotter O sector
while they can be negligible for the colder M sector,
which could provide reasonable differences between the two worlds
in case the exchange process is not too far from equilibrium.
The possible marriage between dark matter and the leptobaryogenesis
mechanism is certainly an attractive feature of our scheme.

The fast reactions relevant for the O-sector are the $\Delta L=1$
one $ l_i \phi \to \bar \lpr_k \barphpr $, and the $\Delta L=2$
ones like $l\phi \to \barl\barphi$, $ll\to \phi\phi$ etc.
Their total rates are correspondingly 
\beqn{rates} 
\Ga_{1} = \frac{Q_{1}n_{\rm eq} }{8\pi M^2};  ~~~~ 
Q_1={\rm Tr}(D^\dagger D) =
{\rm Tr}[(Y^{\prime\dagger}Y^\prime)^\ast G^{-1}
(Y^\dagger Y) G^{-1} ] ,  \nonumber \\
\Ga_{2} = \frac{3Q_{1}n_{\rm eq}}{4\pi M^2};  ~~~~ 
Q_2={\rm Tr}(A^\dagger A) =
{\rm Tr}[(Y^{\dagger}Y)^\ast G^{-1}(Y^\dagger Y) G^{-1}] , 
\eeqn
where $n_{eq}\simeq (1.2/\pi^2)T^3$ is an equilibrium density  
per (bosonic) degree of freedom, and the sum is taken over 
all flavour and isospin indices of initial and final states.  
It is essential that these processes stay out of equilibrium,  
which means that their rates should not exceed much 
the Hubble parameter $H = 1.66 \, g_\ast^{1/2} T^2 / M_{Pl} $
for temperatures $T \leq T_R$, where 
$g_\ast$ is the effective number of particle degrees of freedom, 
namely $g_\ast \simeq 100$ in the SM. 
%and $g_\ast\simeq 200$ for its supersymmetric extension.  
In other words, the dimensionless parameters 
\beqn{K12} 
k_1 = \left({{\Gamma_1} \over {H}}\right)_{T=T_R} 
\simeq 1.5 \times 10^{-3}\, {{Q_1 T_R M_{Pl}} \over 
{g_\ast^{1/2}M^2}}  \,  \nonumber \\
k_2 = \left({{\Gamma_2} \over {H}}\right)_{T=T_R} 
\simeq 9 \times 10^{-3}\, {{Q_2 T_R M_{Pl}} \over 
{g_\ast^{1/2}M^2}}  \, 
\eeqn
should not be much larger than 1.

\begin{figure}[t]
  \begin{center}
    \leavevmode
    \epsfxsize = 7cm
    \epsffile{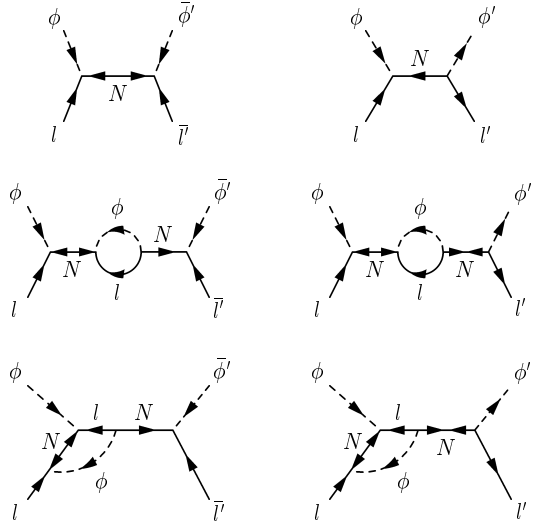}
  \end{center}
\caption{\small Tree-level and one-loop
diagrams contributing to the CP-asymmetries in
$l \phi \to \barlpr \barphpr$ (left column) and  
$l \phi \to \lpr \phpr$ (right column).}
\label{fig1}
\end{figure}

\begin{figure}[t]
  \begin{center}
    \leavevmode
    \epsfxsize = 7cm
    \epsffile{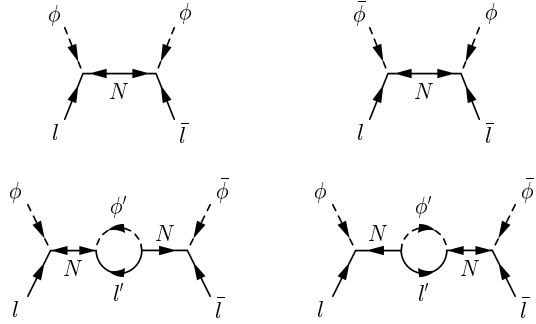}
  \end{center}
\caption{\small Tree-level and one-loop diagrams contributing  
to the CP-asymmetry of $l \phi \to \barl \barphi$.
The external leg labels identify the initial and final state particles.}
\label{fig2}
\end{figure}

Let us now turn to CP-violation.
In $\Delta L=1$ processes the CP-odd lepton number asymmetry
emerges from the interference between the tree-level and
one-loop diagrams of fig.~\ref{fig1}. 
However, CP-violation takes also place in $\Delta L=2$ processes 
(see fig.\ \ref{fig2}).
This is a consequence of the very existence of the mirror sector,
namely, it comes from the contribution of the mirror particles 
to the one-loop diagrams of fig.\ \ref{fig2}. 
The direct calculation gives:\footnote{
It is interesting to note that the tree-level amplitude for the 
dominant channel $l\phi\to \barlpr\barphpr$ goes as $ 1/M$ and the
radiative corrections as $ 1/M^3$.
For the channel $l\phi\to \lpr\phpr$ instead, both tree-level and
one-loop amplitudes go as $ 1/M^2$.
As a result,  the cross section CP asymmetries are comparable 
for both $l\phi\to \barlpr\barphpr$ and
$l\phi\to \lpr\phpr$ channels. } 
\begin{eqnarray}\label{CP}
&&
\sigma (l\phi\to \barlpr\barphpr) -
\sigma(\barl\barphi \to \lpr\phpr) =
(- \Delta\sigma  - \Delta\sigma' ) /2
\, ,  \nonumber \\
&&
\sigma (l\phi\to \lpr\phpr) -
\sigma(\barl\barphi \to \barlpr\barphpr) =
( -\Delta\sigma + \Delta\sigma' )/2
\, ,   \nonumber  \\
&&
\sigma (l\phi\to \barl\barphi) -   
\sigma(\barl\barphi \to l\phi) = \Delta\sigma \, ;  \\
&&
\Delta\sigma = {{3J\, S} \over {32\pi^2 M^4}} \, , ~~~~~
\Delta\sigma' = {{3J'\, S} \over {32\pi^2 M^4}} \, ,
\end{eqnarray}
where $S$ is the c.m.\ energy square,  
$J= {\rm Im\, Tr} [ (Y^{\dagger}Y)^\ast G^{-1}
(Y^{\prime\dagger}Y^\prime) G^{-2} 
(Y^\dagger Y) G^{-1}] $ is the CP-violation parameter and
$J'$ is obtained from $J$ by exchanging
$Y$ with $Y^\prime$. The contributions yielding asymmetries 
$\mp \Delta \sigma'$ respectively for $l\phi\to \barlpr\barphpr$ and
$l\phi\to \lpr\phpr$ channels emerge from the diagrams with $\lpr\phpr$ 
inside the loops, not shown in fig.\ \ref{fig1}.

%  FIGURA  %
% --------------- %

%  FIGURA  %
% --------------- %

This is in  perfect agreement with CPT invariance that requires
that the total cross sections for particle and anti-particle
scatterings are equal to each other:
$\sigma(l\phi \to X) = \sigma(\barl\barphi \to X)$.
Indeed, taking into account that
$\sigma(l\phi \to l\phi) = \sigma(\barl\barphi \to \barl\barphi)$
by CPT, we see that CP asymmetries in the $\Delta L=1$ and
$\Delta L=2$ processes should be related as
\be{CPT}
\sigma(l\phi \to X^\prime) - \sigma(\barl\barphi \to X^\prime) =
- [ \sigma (l\phi\to \barl\barphi) -
\sigma(\barl\barphi \to l\phi) ] = - \Delta \sigma   \;,
\ee
where $X^\prime$ are the mirror sector final states,
$\barlpr\barphpr$ and $\lpr\phpr$.
That is, the  $\Delta L=1$ and $\Delta L=2$ reactions have
CP asymmetries with equal intensities but opposite signs.

But, as $L$ varies in each case by a different amount, a net
lepton number decrease is produced, or better, a net increase
of $B-L$ $ \propto \Delta\sigma$
(recall that the lepton number $L$ is violated by the sphaleron
processes, while $B-L$ is changed solely by the above processes).

As far as we assume that the mirror sector is cooler and thus
depleted of particles, the only relevant reactions are the ones with
ordinary particles in the initial state. Hence,  the evolution of
the $B-L$ number density is determined by the CP asymmetries   
shown in eqs.~(\ref{CP}) and obeys the equation
\be{L-eq-2}
{{ d n_{B-L} } \over {dt}} + 3H n_{B-L} + \Gamma n_{B-L} =
\frac34 \Delta\sigma \, n_{\rm eq}^2 = 
1.8 \times 10^{-3}\, \frac{T^8}{M^4}  \;,
\ee
where $ \Gamma = \Gamma_1 + \Gamma_2$ 
% $( Q_1 + 6Q_2 ) n_{\rm eq} / 8\pi M^2 $ 
is the total rate of the $\Delta L=1$ and $\Delta L=2$ reactions, 
and for the CP asymmetric cross section $\Delta\sigma$ we take
the thermal average c.m.\ energy square $S \simeq 17\, T^2$. 

It is instructive to first solve this equation in the limit 
$k_{1,2} \ll 1$, when the out-of-equilibrium conditions 
are strongly satisfied and thus the term $\Gamma n_{B-L}$ 
can be neglected. Integrating this equation we obtain 
for the final $B-L$ asymmetry of the Universe, 
$B-L = n_{B-L}/s$, 
%where $s=(2\pi^2/45)g_\ast T^3$ is the entropy density, 
the following expression:\footnote{ 
Observe that the magnitude of the produced
$B-L$ strongly depends on the temperature, namely,
larger $B-L$ should be produced in the patches where the plasma
is hotter. 
In the cosmological context, this would lead to a situation
where, apart from the adiabatic density/temperature perturbations,
there also emerge correlated isocurvature fluctuations with variable
$B$ and $L$ which could be tested with the future data on the
CMB anisotropies and large scale structure.}
\be{BL}
(B-L)_0 \approx 2 \times 10^{-3} \,
{{J\, M_{Pl} T_R^3} \over {g_\ast^{3/2} M^4 }} . 
%= \frac{20JT_R}{g_\ast^{1/2}(Q_1+Q_2)^2M_{Pl}} \, k^2    
%\approx 10^{-10} \, \frac{J}{(Q_1+Q_2)^2} T_9 k^2  
%\over {M_{12}^{4}}} \, ,
\ee
It is interesting to note that 3/5 of this value 
is accumulated at temperatures $T>T_R$ 
and it corresponds to the amount of $B-L$ produced 
when the inflaton field started to decay and the
particle thermal bath was produced 
%before the reheating temperature is established.
(Recall that the maximal temperature at the reheating period is
usually larger than $T_R$.)
In this epoch the Universe was still dominated by the inflaton
oscillations and therefore it expanded as $a\propto t^{2/3}$ 
while the entropy of the Universe was growing as $t^{5/4}$.
The other 2/5 of (\ref{BL}) is produced at $T<T_R$, 
radiation dominated era when the Universe expanded as 
$a\propto t^{1/2}$ with conserved entropy (neglecting 
the small entropy leaking from the O- to the M-sector).   

This result (\ref{BL}) can be recasted as follows
\be{B-L}
(B-L)_0 \approx 
%2 \times 10^{-3} \,{{J\, M_{Pl} T_R^3} \over {g_\ast^{3/2} M^4 }}
\frac{20 J k^2 T_R}{g_\ast^{1/2} Q^2 M_{Pl}}    
\approx 10^{-10} \, \frac{Jk^2 }{Q^2} 
\left(\frac{T_R}{10^9~ {\rm GeV}}\right)   
%\over {M_{12}^{4}}} \, ,
\ee
where $Q^2=Q_1^2+Q_2^2$, $k=k_1+k_2$ and 
we have taken again $g_\ast \approx 100$.
This shows that for Yukawa constants spread e.g.\ in the range
$ 0.1-1 $, one can achieve $ B-L = {\cal O}(10^{-10}) $
for a reheating temperature as low as $ T_R\sim 10^9 $ GeV.
Interestingly, this coincidence with the upper bound from the
thermal gravitino production, $ T_R < 4\times 10^9 $ GeV or so
\cite{Ellis}, indicates that our scenario could also work
in the context of supersymmetric theories.

Let us solve now eq. (\ref{L-eq-2}) without assuming 
$\Ga \ll H$. In this case we obtain \cite{BBC}: 
\be{BLO} 
B-L = (B-L)_0 \cdot D(k) \;,
\ee
where $(B-L)_0$ is the solution of eq.~(\ref{L-eq-2}) in the limit 
$\Ga \ll H$, given by expressions (\ref{BL}) or (\ref{B-L}), 
and the depletion factor $D(k)$ is given by
%with $k=\Gamma/H = k_1 + k_2$, and the functions
\be{Dk}
D(k) = \frac35\, e^{-k} F(k) + \frac25\, G(k) 
\ee
where 
\beqn{Fy}
&& 
F(k) = \frac{1}{4 k^4} \left[(2k -1)^3 + 6k-5 +6 e^{-2k}\right] , 
\nonumber \\
&& 
G(k) = \frac{3}{k^3} \left[ 2 -(k^2 + 2k +2) e^{-k} \right] . 
\eeqn
These two terms in $D(k)$ 
correspond to the integration of (\ref{L-eq-2}) respectively
in the epochs before and after reheating 
($T > T_R$ and $T < T_R$).
Obviously, for $ k \ll 1 $ the depletion factor 
$ D(k) \to 1 $ and thus we recover the result as in (\ref{BL}) or
(\ref{B-L}): $ B-L = (B-L)_0 $.
However, for large $k$ the depletion can be reasonable,
e.g. for $k=1,2$ we have respectively $D(k) = 0.34, 0.1$.

Now, let us discuss how the mechanism considered above produces
%not only the lepton number in the ordinary sector,
also the baryon prime asymmetry in the mirror sector.
The amount of this asymmetry will depend on the CP-violation
parameter $ J^\prime = {\rm Im\, Tr}
[ (Y^\dagger Y) G^{-2} (Y^{\prime\dagger} Y^\prime)
Y^{-1} (Y^{\prime\dagger} Y^\prime)^\ast G^{-1}] $ 
that replaces $J$ in $ \Delta \sigma' $ of eqs.\ (\ref{CP}). 
The mirror P parity under the exchange $ \phi \to \phi^{\prime\dagger} $,
$ l \to \bar{l}^\prime $, etc.,
implies that the Yukawa couplings are essentially the same in
both sectors, $ Y^\prime= Y^\ast$.
Therefore, in this case also the CP-violation parameters are
the same, $ J^\prime = J $.\footnote{It is interesting to remark 
that this mechanism needs the left-right parity $P$ rather than the 
direct doubling  one $D$. One can easily see that the 
latter requires $Y^\prime= Y$, and so the CP-violating 
parameters $J$ and $J'$ are both vanishing.}  
Therefore, one naively expects that
$ n'_{B-L} = n_{B-L} $ and the mirror baryon density 
should be equal to the ordinary one,
$ \Omega'_{b} = \Omega_{b} $.

However, now we show that if the $ \Delta L = 1 $ and
$ \Delta L = 2 $ processes are not very far from equilibrium,
i.e. $ k \sim 1 $, the mirror baryon density should
be bigger than the ordinary one.
Indeed, the evolution of the mirror $B-L$  number density, 
$n'_{B-L}$, obeys the equation
\be{L-eq-3}
{{ d n'_{B-L} } \over {dt}} + 3H n'_{B-L} + \Gamma' n'_{\rm B-L} =
  {3 \over 4} \Delta\sigma' \, n_{\rm eq}^2   \; ,
\ee
where now $ \Gamma' = ( Q_1 + 6Q_2 ) n'_{\rm eq} / 8\pi M^2 $
is the total reaction rate of the
$ \Delta L' = 1 $ and $ \Delta L' = 2 $ processes in the mirror sector,
and $ n'_{\rm eq} = (1.2 / \pi^2) T^{\prime 3} = x^3 n_{\rm eq} $
is the equilibrium number density per degree of freedom in the 
mirror sector. Therefore $k'=\Gamma'/H = x^3 k $,
and thus for the mirror sector we have
$ (B-L)' = (B-L)_0 D(kx^3)$,  
where the depletion can be irrelevant if $kx^3 \ll 1$.  

Now taking into the account that in both sectors the
$B-L$ densities are reprocessed into the baryon number densities
by the same sphaleron processes,
we have $B = a(B-L)$ and $B' = a(B-L)'$,
with coefficients $a$ equal for both sectors.
Therefore, we see that the cosmological densities of the ordinary 
and mirror baryons should be related as
\be{omegabp}
\beta = \frac{\Om'_b}{\Om_b} \approx \frac{1}{D(k)}
%\Omega'_{\rm b} = {D(2K') \over D(2K)} \: \Omega_{\rm b} \;.
\ee
If $k \ll 1$,
depletion factors in both sectors are $ D \approx D' \approx 1 $
and thus we have that the mirror and ordinary baryons have the
same densities,
$\Omega'_{\rm b} \approx \Omega_{\rm b}$.
In this case mirror baryons are not enough to explain all dark 
matter and one has to invoke also some other kind of dark matter,
presumably cold dark matter.

However, if $k \sim 1$, then we would have
$ \Omega'_{\rm b} > \Omega_{\rm b} $,
and thus all dark matter of the Universe could be in the form
of mirror baryons.
Namely, for $k = 1.5$ we would have from eq.~(\ref{omegabp})
that $\Omega'_{b}/\Omega_{b} \approx 5$,
%and hence for $ \Omega_{\rm b} \approx 0.05 $ we have
%$ \Omega'_{\rm b} \approx 0.25 $, 
which is about the best fit relation between the ordinary 
and dark matter densities. 

On the other hand, eq. (\ref{B-L}) shows that $k\sim 1$ 
is indeed preferred for explaining the observed magnitude 
of the baryon asymmetry. For $k\ll 1$ the result could be too 
small, since $(B-L)_0\propto k^2$ fastly goes to zero. 
%if $k\ll 1$.  

One could question, whether the two sectors would not equilibrate 
their temperatures if $k\sim 1$. 
%and thus whether the unbalanced leaking between two
As far as the mirror sector includes the gauge couplings  
which are the same as the standard ones, 
the mirror particles should be
thermalized at a temperature $ T^\prime$.
Once $ k_1 \leq 1 $, $T^\prime$ will remain smaller than the 
parallel temperature $T$ of the ordinary system, and so 
the presence of the out-of-equilibrium hidden sector does
not affect much the Big Bang nucleosynthesis epoch.

Indeed, if the two sectors had different temperatures at 
reheating, then they evolve independently during
the expansion of the Universe and approach the nucleosynthesis era 
with different temperatures. For $ k_1\leq 1 $, the energy 
density transferred to the mirror sector will be crudely
$ \rho^\prime \approx (8 k_1/g_\ast)\rho$ \cite{BB-PRL},
where $ g_\ast(\approx 100) $ is attained to the leptogenesis epoch.
Thus, translating this to the BBN limits, 
this corresponds to a contribution equivalent to an 
effective number of extra light neutrinos 
$\Delta N_\nu \approx k/14$.

\section{Mirror baryons as dark matter}

We have shown that mirror baryons could provide
a significant contribution to the energy density
of the Universe and thus they could constitute a  
relevant component of dark matter.
An immediate question arises:
how the mirror baryon dark matter (MBDM) behaves
and what are the differences from the more familiar dark
matter candidates as the cold dark matter (CDM),
the hot dark matter (HDM) etc.
In this section we briefly address the
possible observational consequences of
such a cosmological scenario.

In the most general context, the present energy
density contains a relativistic (radiation) component
$\Omega_r$, a non-relativistic (matter) component
$\Omega_m$ and the vacuum energy
density $\Omega_\Lambda$ (cosmological term).
According to the inflationary paradigm the Universe
should be almost flat,
$\Omega_0=\Omega_m + \Omega_r + \Omega_\Lambda \approx 1$,
which agrees well with the recent results
on the CMBR anisotropy and large scale power spectrum.
%(in units of the critical density
%$\rho_c = 3H_0^2/8\pi G_N$, where

The Hubble parameter is known to be
$H_0 = 100 h$ km s$^{-1}$ Mpc$^{-1}$ with
$h \approx 0.7$, and for redshifts of cosmological
relevance, $1+z = T/T_0 \gg 1$, it becomes
\be{H}
H(z)= H_0 \left[\Omega_{r}(1+z)^4
+ \Omega_{m} (1+z)^3 + \Omega_\La \right]^{1/2}  .
\ee
In the context of our model, the relativistic fraction
is represented by the ordinary and mirror
photons and neutrinos,
$\Omega_rh^2=4.2\times 10^{-5}(1+x^4)$, and the 
contribution of the mirror species is negligible
in view of the BBN constraint $x< 0.6$.
As for the non-relativistic component,
it contains the O-baryon fraction $\Omega_b$ and
%$\omega_b=\Omega_bh^2 = 0.02$ and M-baryons with
the M-baryon fraction $\Omega'_b = \beta\Omega_b$, 
while the other types of dark matter, e.g. the CDM,
could also be present. Therefore, in general, 
$\Omega_m=\Omega_b +\Omega'_b+\Omega_{\rm cdm}$.\footnote{
In the context of supersymmetry,
the CDM component could exist in the form of
the lightest supersymmetric particle (LSP).
It is interesting to remark that the mass fractions
of the ordinary and mirror LSP are related as
$\Omega'_{\rm LSP} \simeq x\Omega_{\rm LSP}$.
%In addition, a significant HDM component $\Omega_\nu$
%could be due to neutrinos with order eV mass.
The contribution of the mirror neutrinos
scales as $\Omega'_\nu = x^3 \Omega_\nu$ and thus
it is also irrelevant.
}

The important moments for the structure formation
are related to the matter-radiation equality (MRE) epoch
and to the plasma recombination and matter-radiation
decoupling (MRD) epochs.

The MRE occurs at the redshift 
\be {z-eq}
1+z_{\rm eq}= \frac{\Omega_m}{\Omega_r} \approx
%4.02\cdot 10^4 \frac{\omega_{m}}{(1+0.227\,N_\nu)(1+x^4)}
 2.4\cdot 10^4 \frac{\omega_{m}}{1+x^4} 
% = 4800 \times (\Omega_{m}h^2)_{0.2}
\ee
where we denote $\omega_m = \Omega_{m}h^2$. 
Therefore, for $x\ll 1$ it is not altered by the additional 
relativistic component of the M-sector. 

The radiation decouples from matter after almost all of
electrons and protons recombine into neutral hydrogen
and the free electron number density  sharply diminishes,
so that the photon-electron scattering rate
%$\Gamma_\gamma=n_{e}\sigma_{T}=X_{e}\eta n_{\gamma} \sigma_{T}$
drops below the Hubble expansion rate. 
In the ordinary Universe the MRD takes place
in the matter domination period, at the temperature
$T_{\rm dec} \simeq 0.26$ eV, which corresponds to the redshift
$1+z_{\rm dec}=T_{\rm dec}/T_0 \simeq 1100$.

\begin{figure}
\centerline{\psfig{file=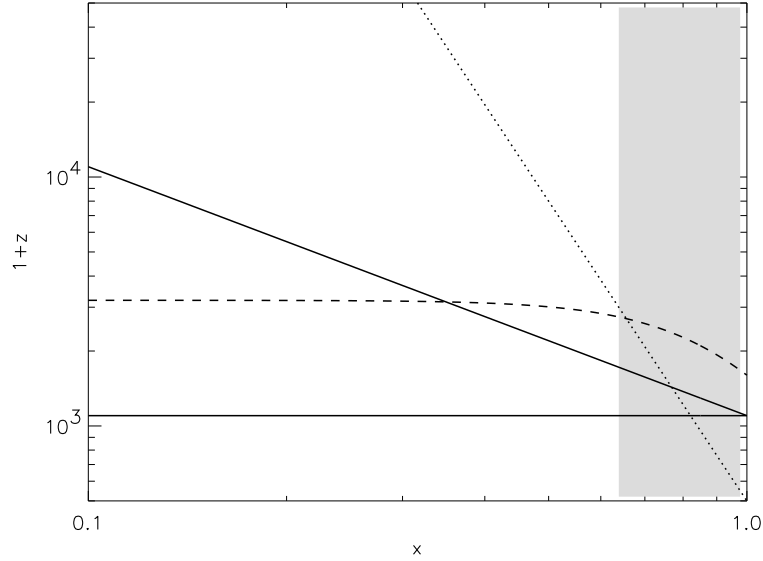,width=10cm}}
%\centerline{\psfig{file=zu-scales.eps,width=10cm}}
%\vspace*{-128pt}
\caption{
The M-photon decoupling redshift $1+z_{dec}'$ as a function of $x$ 
(thick solid). The horizontal thin solid line marks the ordinary 
photon decoupling redshift $1+z_{\rm dec} = 1100$. 
We also show the matter-radiation equality redshift
$1+z_{\rm eq}$ (dash) and the mirror Jeans-horizon mass equality
redshift $1+z'_c$ (dash-dot) for the case $\omega_m =0.135$. 
The shaded area $x> 0.64$ is excluded by the BBN limits. 
}
\label{fig3} 
\end{figure}

The MRD temperature in the M-sector $T'_{\rm dec}$
can be calculated following the same lines as in
the ordinary one \cite{BCV}.
Due to the fact that in either case the
photon decoupling occurs when the exponential factor
in Saha equations becomes very small,
we have $T'_{\rm dec} \simeq T_{\rm dec}$,
up to small logarithmic corrections related to
$B'$ different from $B$. Hence
\be{z'_dec}
1+z'_{\rm dec} \simeq x^{-1} (1+z_{\rm dec})
\simeq 1100\, x^{-1}
\ee
so that the MRD in the M-sector occurs earlier
than in the ordinary one. Moreover, for $x$ less than
$x_{\rm eq}=0.045\omega_m^{-1}\simeq 0.3$,
%$x < x_{\rm eq}=0.045(\Omega_m h^2)^{-1}$,
the mirror photons would decouple
yet during the radiation dominated period
(see Fig. \ref{fig3}).

%%%%%%%%%%%%%%%%%%%%%%%%%%%%%%%%%%%%%%
%%%%%%%%%%%%%%%%%%%%%%%%%%%%%%%%%%%%%%%%%%%%%%%%%%%%%%%% 

Let us now discuss the cosmological evolution of the MBDM.
The relevant length scale for the gravitational
instabilities is characterized by the mirror Jeans scale
$\lambda'_J \simeq v'_s (\pi/G\rho)^{1/2}$,
where $\rho(z)$ is the matter density at a given redshift
$z$ and $v'_s(z)$ is the sound speed in the M-plasma.
The latter contains more baryons and less photons than the
ordinary one, $\rho'_b=\beta\rho_b$ and
$\rho'_\gamma = x^4\rho_\gamma$.
Let us consider for simplicity the case
when dark matter of the Universe is entirely due to
M-baryons, $\Omega_m\simeq\Omega'_b$. Then we have:
\be{sound}
v'_s(z) \simeq \frac{c}{\sqrt3}
\left(1+ \frac{3\rho'_b}{4\rho'_\gamma}\right)^{-1/2} 
\approx \frac{c}{\sqrt3}
\left[ 1 +\frac34\left(1+x^{-4}\right)
\frac{1+z_{\rm eq}}{1+z}\right]^{-1/2} .   
\ee
Hence, for redshifts 
of cosmological relevance, $z\sim z_{\rm eq}$,
we have $v'_s \sim 2x^2 c/3 \ll c/\sqrt{3}$,
quite in contrast with the ordinary world,
where $v_s \approx c/\sqrt{3}$ practically
until the photon decoupling, $z=1100$.

The M-baryon Jeans mass
$M'_J =\frac{\pi}{6} \rho_m \lambda'^3_J$ reaches the
maximal value at $z=z'_{\rm dec}\simeq 1100/x$,
$M'_J(z'_{dec}) \simeq 2.4 \cdot 10^{16}\times
x^6 [1+(x_{\rm eq}/x)]^{-3/2}
\omega_m^{-2} ~ M_\odot$.   
Notice, however, that $M'_J$ becomes smaller than the
Hubble horizon mass $M_H = \frac{\pi}{6} \rho H^{-3}$
starting from a redshift
$z_c= 3750 x^{-4} \omega_m $, which is
about $z_{\rm eq}$ for $x=0.64$, but
it sharply increases for smaller values of $x$
(see Fig. \ref{fig3}).
So, the density perturbation scales which enter
the horizon at $z \sim z_{\rm eq}$ have mass larger
than $M'_J$ and thus undergo uninterrupted linear growth
immediately after $t=t_{\rm eq}$.
The smaller scales for which $M'_J > M_H$
instead would first oscillate.
Therefore, the large scale structure
formation is not delayed even if the mirror MRD epoch
did not occur yet, i.e. even if $x> x_{\rm eq}$.
The density fluctuations start to grow in the M-matter
and the visible baryons are involved later, when after 
being recombined they fall into the potential whells 
of developed mirror structures.

Another important feature of the MBDM scenario is that the
M-baryon density fluctuations should undergo  
strong collisional damping around the time of  
M-recombination.
The photon diffusion from the overdense to underdense
regions induce a dragging of charged particles
and wash out the perturbations at scales smaller than the
mirror Silk scale $\lambda'_S \simeq
3\times f(x)\omega_m^{-3/4}$ Mpc,
where $f(x)=x^{5/4}$ for $x > x_{\rm eq}$,
and $f(x) = (x/x_{\rm eq})^{3/2} x_{\rm eq}^{5/4}$
for $x < x_{\rm eq}$.

Thus, the density perturbation scales which can undergo 
the linear growth after the MRE epoch are limited by the
length $\lambda'_S$.
This could help in avoiding the excess of small scales
(of few Mpc) in the power spectrum without
tilting the spectral index.
The smallest perturbations that survive the
Silk damping will have the mass
$M'_S \sim f^3(x) \omega_m^{-5/4} \times 10^{12}~ M_\odot $. 
%which should be less than $10^{13} ~ M_\odot$
%in view of the BBN bound $x <0.64$.
Interestingly, for $x\sim x_{\rm eq}$ we have
$M'_S \sim 10^{11}~M_\odot$, a typical galaxy mass.   
To some extend, the cutoff effect is
analogous to the free streaming damping in the case of
warm dark matter (WDM), but there are important   
differences. The point is that like usual baryons,
the MBDM should show acoustic oscillations 
whith an impact on the large scale power spectrum.

%%%%%%%%%%%%%%%%%%%%%%%%%%%%%%%%%%%%%%%%%%%%%%%%

%  FIGURA  %
\begin{figure}[h]
  \begin{center}
    \leavevmode
    \epsfxsize = 11cm
    \epsffile{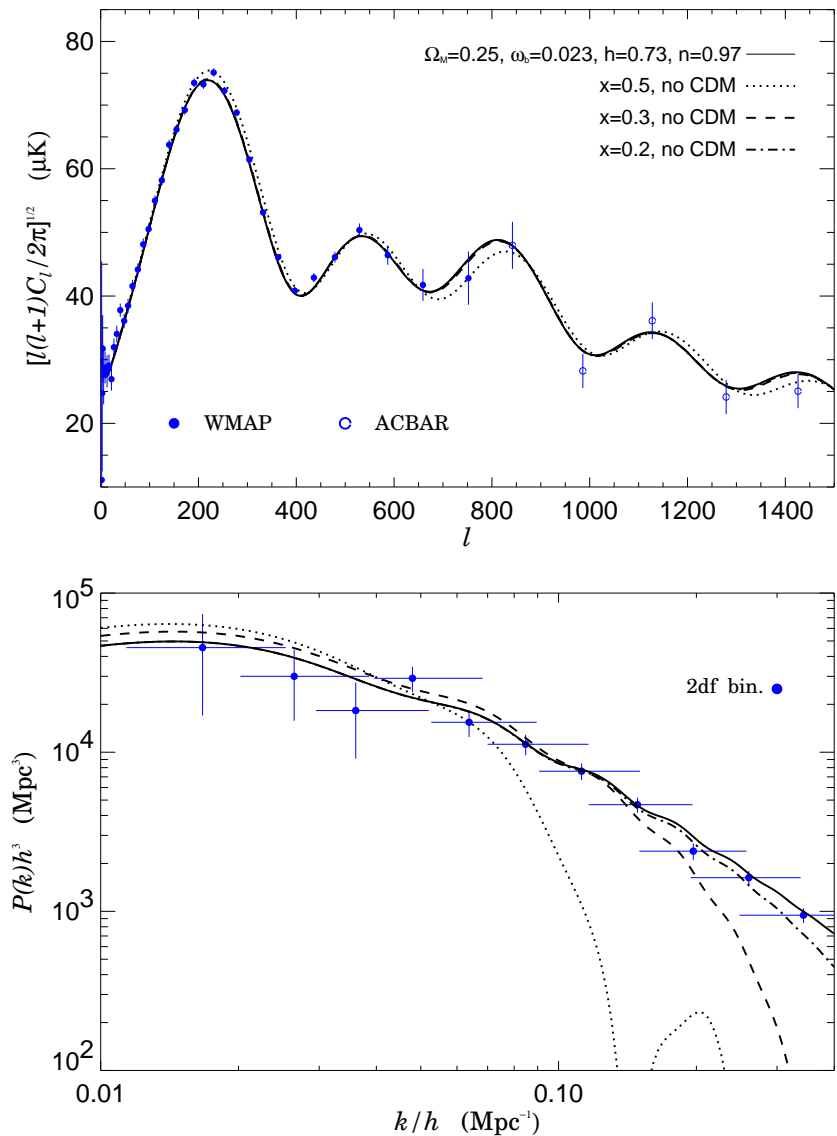}
%    \epsffile{zu-cmblss2.eps}
  \end{center}
\caption{\small The CMBR power spectrum (upper panel) 
and the large scale power spectrum (lower panel) for a 
"concordance" set of cosmological parameters. 
The solid curves correspond to the flat $\Lambda$CDM model, 
while dot, dash and dash-dot curves correspond to 
the situation when the CDM component is completely 
substituted by the MBDM for different values of $x$. 
}
\label{fig4}
\end{figure}
% --------------- %

In addition, the MBDM oscillations transmitted
via gravity to the ordinary baryons,
could cause observable anomalies in the CMB
angular power spectrum for $l$'s larger than 200.  
This effect can be observed only if the M-baryon Jeans
scale $\lambda'_J$ is larger than the Silk scale  
of ordinary baryons, 
%$\lambda_S \simeq 3\beta^{1/2} (\Omega_m h^2)_{0.2}^{-3/4}$ Mpc,
which sets a principal cutoff for CMB oscillations
around $l\sim 1200$.
As we have seen above, this would require enough large
values of $x$, near the BBN upper bound $x \simeq 0.6$ or so.

If the dark matter is entirely built up by 
mirror baryons, large values of $x$ are 
excluded by the observational data. For the sake 
of demostration, on Fig. \ref{fig4} we show the 
CMBR and LSS power spectra for different values of $x$. 
We see that for $x > 0.3$ the matter power spectrum 
shows a strong deviation from the experimental data. 
This is due to Silk damping effects which suppress 
the small scale power too early, already for 
$k/h\sim 0.2$. However, the values $x<0.3$ are  
compatible with the observational data.

This has a simple explanation. 
Clearly, for small $x$ the M-matter recombines
before the MRE moment, and thus it should rather manifest
as the CDM as far as the large scale structure is concerned.
However, there still can be a crucial difference at
smaller scales which already went non-linear, like galaxies. 
Then one can question whether the MBDM distribution
in halos can be different from that of the CDM?
Namely, simulations show that the CDM forms triaxial
halos with a density profile too clumped towards the
center, and overproduce the small substructures within
the halo. As for the MBDM, it constitutes a sort of
collisional dark matter and thus potentially could avoide
these problems, at least clearly the one related with
the excess of small substructures.

As far as the MBDM constitutes a dissipative dark matter 
like the usual baryons, one would question how it 
can provide extended halos instead of being clumped 
into the galaxy as usual baryons do. 
However, one has to take into account the possibility
that during the galaxy evolution
the bulk of the M-baryons could fastly fragment
into the stars.
%and only small part of them could be in gaseous state,   
A difficult question to address here
is related to the star formation in the M-sector,
also taking into account that its temperature/density 
conditions and chemical contents
are much different from the ordinary ones.
In any case, the fast star formation would
extinct the mirror gas and thus
could avoide the M-baryons to form disk galaxies. 
The M-protogalaxy, which at a certain moment before disk formation
essentially becomes a collisionless system of the  
mirror stars, could maintain a typical elliptical structure.
In other words, we speculate on the possibility
that the M-baryons form mainly elliptical 
galaxies.\footnote{For a comparison, in the ordinary world
the number of spiral and elliptic galaxies are 
comparable. Remarkably, the latter contain old stars, 
very little dust and show low activity of star formation.  
}
Certainly, in this consideration also the galaxy merging
process should be taken into account.
As for the O-matter, within the dark M-matter halo it should
typically show up as an observable elliptic or spiral galaxy,
but some anomalous cases can also be possible,
like certain types of irregular galaxies or even
dark galaxies dominantly made of M-baryons.

%  FIGURA  %
\begin{figure}[h]
  \begin{center}
    \leavevmode
    \epsfxsize = 11cm
    \epsffile{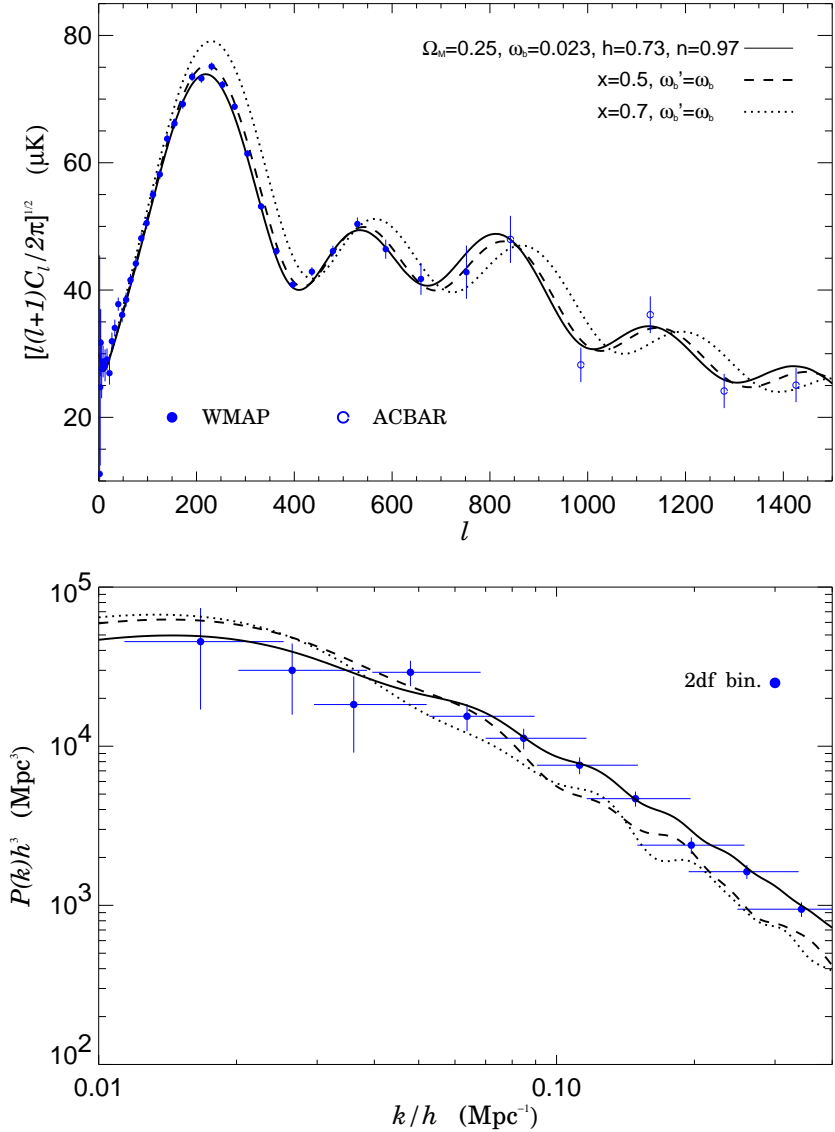}
%    \epsffile{zu-cmblss1.eps}
  \end{center}
\caption{\small The same as on Fig. 4, however for the mixed 
CDM+MBDM scenario for large values of $x$. The ordinary and 
mirror baryon densities are taken equal, $\Om_b'=\Om_B$, 
and the rest of matter density is attained to the SDM component.}
\label{fig5}
\end{figure}
% --------------- %

Another tempting issue is whether the M-matter itself
could help in producing big central black holes,
with masses up to $\sim 10^9~ M_\odot$, which are thought
to be the main engines of active galactic nuclei.  

Another possibility can also be considered when 
dark matter in galaxies and clusters contain mixed CDM and
MBDM components, $\Om_d=\Om'_b+\Om_{cdm}$. 
e.g.  one can exploit the case when mirror baryons 
constitute the same fraction of matter as the ordinary ones, 
$\Om'_b=\Om_b$, a situation which emerges naturally in the  
leptogenesis mechanism of sect. 4.3 in the case of small $k$. 

In this case the most interesting and falsificable predictions 
are related to the large $x$ regime. On Fig. \ref{fig5} we show 
the results for the CMBR and LSS power spectra. We see that 
too large values of $x$ are excluded by the CMBR anisotropies, 
but e.g. $x \leq 0.5$ can still be compatible with the data. 

The detailed analysis of this effect will be given elsewhere 
\cite{BCCV}. 
In our opinion, in case of large $x$ the effects on
the CMBR and LSS  can provide direct tests for the MBDM 
and can be falsified by    
the next observations with higher sensitivity.

In the galactic halo
(provided that it is an elliptical mirror galaxy)
the mirror stars should be observed as
Machos in gravitational microlensing  \cite{BDM,Macho}.
Leaving aside the difficult question of the initial
stellar mass function, one can remark that once 
the mirror stars could be very old
and evolve faster than the ordinary ones,
it is suggestive to think that most   
of the massive ones, with mass above the
Chandrasekhar limit $M_{\rm Ch} \simeq 1.5 ~ M_\odot$, 
have already ended up as supernovae, so that only the  
lighter ones 
%like white dwarfs or neutron stars
remain as the microlensing objects.
The recent data indicate the average mass of
Machos around $M\simeq 0.5 ~M_\odot$, which is difficult
to explain in terms of the brown dwarves with masses 
below the hydrogen ignition limit $M < 0.1 M_{\odot}$  
or other baryonic objects \cite{Freese}.
Perhaps, this is the observational evidence
of mirror matter?

It is also possible that in the galactic halo
some fraction of mirror stars exists in the form  
of compact substructures like globular or open clusters.
In this case, for a significant statistics, one could  
observe interesting time and angular correlations
between the microlensing events.

The explosions of mirror supernovae in our galaxy cannot be directly
seen by an ordinary observer. 
However, it should be observed in terms of gravitational waves.
In addition, if the M- and O-neutrinos are mixed \cite{FV,BM},
it can lead to an observable neutrino signal, and could 
be also accompanied by a weak gamma ray burst \cite{GRB}.

\section{Conclusions and outlook}

We have discussed cosmological implications of the   
parallel mirror world with the same microphysics
as the ordinary one, but having smaller temperature,
$T'< T$, with the limit on $x=T'/T<0.6$  set by  
the BBN constraints.
Therefore, the M-sector contains less relativistic
matter (photons and neutrinos) than the O-sector,
$\Omega'_r \ll \Omega_r$.
On the other hand, in the context of certain 
baryogenesis scenarios, the condition
$T'<T$ yields that the mirror sector should produce a
larger baryon asymmetry than the observable one, 
$B'>B$.
So, in the relativistic expansion epoch the cosmological
energy density is dominated by the ordinary component,
while the mirror one gives a negligible contribution.
However, for the non-relativistic epoch
the complementary situation can occur when
the mirror baryon density is bigger
than the ordinary one, $\Omega'_b > \Omega_b$.
Hence, the MBDM can contribute as dark matter along with 
the CDM or even entirely  constitute it.

Unfortunately, we cannot exchange the information
with the mirror physicists and combine our observations.
(After all, since the two worlds have the same microphysics,
life should be possible also in the mirror sector.)
However, there can be many possibilities to disentangle
the cosmological scenario of two parallel worlds
with the future high precision data concerning  
the large scale structure, CMB anisotropy,
structure of the galaxy halos, gravitational
microlensing, oscillation of neutrinos or other
neutral particles into their mirror partners, etc.

Let us conclude with two quotes of a renowned 
theorist. In 1986 Glashow found a contradiction  
between the estimates of the GUT scale induced kinetic 
mixing term (\ref{FFpr})
and the positronium limits $\eps\leq 4\times 10^{-7}$ 
and concluded that \cite{Glashow86}:  
{\sl "Since these are in evident conflict, the notion of a mirror 
universe with induced electromagnetic couplings of plausible 
(or otherwise detectable) magnitudes is eliminated. The unity 
of physics is again demonstrated when the old positronium workhorse 
can be recalled to exclude an otherwise tenable hypothesis".} 

The situation got another twist within one year, after   
the value $\eps\approx 4 \times 10^{-7}$ appeared to be just  
fine for tackling the mismatch problem of the orthopositronium 
lifetime. However, in 1987 Glashow has fixed that this value  
was in conflict with the BBN limit $\eps < 3\times 10^{-8}$ 
and concluded the following \cite{Glashow87}:  
{\sl "We see immediately that this limit on $\epsilon$ excludes 
mirror matter as an explanation of the positronium lifetime \dots
We also note that the expected range for $\epsilon$ 
$(10^{-3}-10^{-8})$ is also clearly excluded. 
This suggests that the mirror universe, if it exists at all, 
couples only gravitationally to our own. 
If the temperature of the mirror universe is much lower than 
our own, then no nucleosynthesis limit can be placed on the 
mirror universe at all. 
Then it is also likely that the mirror universe would have 
a smaller baryon number as well, and hence would be virtually empty. 
This makes a hypothetical mirror universe undetectable at energies 
below the Planck energy. 
Such a mirror universe can have no influence on the Earth and 
therefore would be useless and therefore does not exist".}   

The main purpose of this paper was to object to this statement. 
The mirror Universe, if it exists at all, would be useful 
and can have an influence if not directly on the Earth, 
but on the formation of galaxies ... and moreover, 
the very existence of matter, both of visible and dark components,  
can be a consequence of baryogenesis via entropy exchange 
between the two worlds. 
The fact that the temperature of the mirror Universe is much 
lower than the one in our own, does not imply that it would have 
a smaller baryon number as well and hence would be virtually 
empty, but it is likely rather the opposite, 
mirror matter could have larger baryon number and 
being more matter-rich, 
it can provide a plausible candidate for dark matter in the 
form of mirror baryons. Currently it seems to be the 
only concept which could naturally explain the coincidence 
between the visible and dark matter densities of the 
Universe. In this view, future experiments for direct 
detection of mirror matter are extremely interesting.

%%%%%%%%%%%%%%%%%%%%%%%%%%%%%%%%%%%%%%%%%%%%%%%%%%%%%%%%%%%%
% Doing Acknowledgement                                              %
%%%%%%%%%%%%%%%%%%%%%%%%%%%%%%%%%%%%%%%%%%%%%%%%%%%%%%%%%%%%

\section*{Acknowledgements}

I would like to thank L. Bento, V. Berezinsky, 
S. Borgani, P. Ciarcelluti, D. Comelli, 
A. Dolgov, A. Doroshkevich, S. Gninenko and F. Villante 
for useful discussions and collaborations.  
The work is partially supported by the MIUR
research grant "Astroparticle Physics".

%%%%%%%%%%%%%%%%%%%%%%%%%%%%%%%%%%%%%%%%%%%%%%%%%%%%%%%%%%%%
% Doing Appendix(ices) - Appendix A & B are shown below    %  
%%%%%%%%%%%%%%%%%%%%%%%%%%%%%%%%%%%%%%%%%%%%%%%%%%%%%%%%%%%%

%\appendix

%\section{HEADING FOR APPENDIX A}

%TYPE TEXT FOR APPENDIX A HERE.

%\section{HEADING FOR APPENDIX B}

%TYPE TEXT FOR APPENDIX B HERE.

%%%%%%%%%%%%%%%%%%%%%%%%%%%%%%%%%%%%%%%%%%%%%%%%%%%%%%%%%%%%
% Doing references:                                        %
%%%%%%%%%%%%%%%%%%%%%%%%%%%%%%%%%%%%%%%%%%%%%%%%%%%%%%%%%%%%

\end{document}